\documentclass[preprint]{aastex}
\usepackage{fullpage}
\usepackage{color}
\usepackage{multirow}

\usepackage{natbib}

\slugcomment{Not to appear in Nonlearned J., 45.}
\shorttitle{Monte Carlo Simulation with Bulk Diffusion}
\shortauthors{Lu et al.}

\begin{document}

\title{The Chemical Evolution from Prestellar to Protostellar Cores: A New Multiphase Model With Bulk Diffusion and Photon Penetration}

\author{Yang Lu}
\affil{Xinjiang Astronomical Observatory, Chinese Academy of Sciences, 150 Science 1-Street, Urumqi 830011, PR China}
\author{Qiang Chang}
\affil{Xinjiang Astronomical Observatory, Chinese Academy of Sciences, 150 Science 1-Street, Urumqi 830011, PR China;Email:changqiang@xao.ac.cn}
\affil{Key Laboratory of Radio Astronomy, Chinese Academy of Sciences, 150 Science 1-Street, Urumqi 830011, PR China}
\author{Yuri Aikawa}
\affil{Department of Astronomy, Graduate School of Science, The University of Tokyo, 7-3-1 Hongo, Bunkyo-ku, Tokyo 113-0033, Japan}

\begin{abstract}

We investigate the chemical evolution of a collapsing core that starts from a hydrostatic core and finally
form a low-mass protostar. 
New multiphase gas-grain models that include 
bulk diffusion and photon penetration are simulated by the macroscopic Monte Carlo method
in order to derive the chemical evolution.
There are two types of species in the ice bulk in the new multiphase models; interstitial species can diffuse and sublime 
at their own sublimation temperatures while
normal species are locked in the ice bulk.
Photodissociation rates of icy species are reduced by the exponential decay of UV flux within the ice mantle.
Two-phase models and basic multiphase models without bulk diffusion and photon penetration 
are also simulated for comparison.
Our physical model for the collapsing core is based on a one-dimensional radiation hydrodynamics model.
Abundant icy radicals are produced at around 10 K
in the new multiphase models. 
Interstitial radicals can diffuse inside ice mantles to form complex organic molecules (COMs) upon warming-up.
Thus, COMs produced by radical recombination at higher temperatures
in the new multiphase models are more than one order of magnitude higher than those in the two-phase 
and basic multiphase models. 
Moreover, COMs produced at around 10 K in the new multiphase models 
are about one order of magnitude higher than those in the two-phase model.
Our model shows a reasonable agreement with observations
toward low-mass protostars. Molecular oxygen abundances predicted by our new multiphase models
agree reasonably well with that found in cometary materials.

\end{abstract}

\section{Introduction}

The chemical evolution of the raw materials inside molecular clouds when new stars are 
born raises many important and interesting questions. Answers to these questions should help us to understand
star formation processes or even the origin of life better. Therefore, much work has been done to address these questions. 
The chemical evolution is coupled with the physical processes of star formation because
chemical reactions are sensitive to physical conditions. 
On the other hand, chemistry in turn affects line cooling which will impact the gas temperatures 
and hence the dynamics of molecular cloud evolution.
Currently, low-mass star formation is considered to proceed as follows. A prestellar core collapses by self gravity.
Initially the collapsing core is almost isothermal, and thus cold ($\sim 10$ K), but the contraction heating 
eventually overwhelms the radiative cooling in the central region, which becomes adiabatic. Because the temperature 
and gas pressure increase, the contraction is decelerated so that a hydrostatic core is formed, which is called 
the first hydrostatic core or the first Larson core \citep{Larson1969}.
When the temperature of the central region increases up to $\sim 2000$ K, the hydrogen molecules 
are dissociated to neutral atoms, which are then ionized to be protons and electrons. These endothermic 
reactions make the first core unstable. The second collapse ends up with the second hydrostatic core, i.e. 
the protostar. The dense gas harboring a protostar is called a protostellar core ~\citep{Andre2000}, 
which continues to infall onto the protostar. Radiation from the protostar and accretion shock heat the protostellar core.
So far, physical models of collapsing cores 
used for astrochemical models simulations vary in degrees 
of complexity from simple 0-D~\citep{Garrod2006} models to more complicated 1-D radiation hydrodynamic (RHD) models~\citep{Aikawa2008}, 
to 2-D models~\citep{Weeren2009}, and 3-D RHD models~\citep{Furuya2012} and finally to full 
3-D radiation-magnetohydrodynamic models~\citep{Hincelin2013, Yoneda2016}.
The results of these studies significantly enhance our knowledge about the molecular evolution that occurs
when molecular cores collapse to form stars.

However, there are still at least two questions that have not been well answered. COMs, 
which are defined to be carbon-containing molecules with at least 6 atoms~\citep{Herbst2009}, have been observed toward many astronomical 
sources including low-mass protostars \citep{Ceccarelli2007}. The first question is how these COMs are formed.
The abundances of some terrestrial COMs such as dimethyl ether (CH$_3$OCH$_3$) predicted by theoretical models 
are often more than one order of magnitude lower than the observed values toward low-mass protostars~\citep{Aikawa2008}. 
The second question is how the volatile species diffuse in the ice mantle and 
desorb into the gas phase in warming-up cores. Laboratory experiments show that not all solid CO can sublime at 20 K~\citep{Collings2003, Sandford1988}.
However, most astrochemical models used for collapsing cores are the simple two-phase models.
In these models, no distinction is made between the active surface layer and the bulk ice, which should be less active,  
so that all solid CO molecules sublime at temperature around 20 K. 
It should be noted that the diffusion and desorption of CO are linked to the question in COMs.
In cloud cores, CO is the major carbon carrier. In some protostellar cores, however, the abundance of sublimated
CO is lower than its canonical value~\citep{Fuente2012}.
So far, the interpretation is either that CO 
is locked in water ice, or a fraction of CO is converted to COMs.
There are more advanced astrochemical models used for collapsing cores than the two-phase model. 
Three-phase models are able to distinguish the active layer from the ice mantle~\citep{Hasegawa1992}; however,
the application of three-phase models to the chemical simulation of collapsing cores has only achieved limited success.
The three-phase models used  by \citet{Furuya2015}, \citet{Taquet2012}, 
\citet{Vasyunin2013} and \citet{Vasyunin2017} assume
completely inert ice mantles, so the volatile species are locked in the bulk ice and cannot desorb until the water ice sublime.
The three-phase model in \citet{Garrod2013}  
introduces the swapping mechanism of species in ice mantles, so that these species are able to diffuse; 
CO molecules are thereby able to diffuse out of the ice mantles and then sublime at temperatures above 20 K. However,
no CO molecules can be trapped for a long period of time inside the ice mantles in the three-phase model 
with swapping mechanism, which contradicts the laboratory experiments~\citep{Garrod2013}. 
      
Improved surface astrochemical models are required to answer the above two questions better.
Theoretical models and laboratory experiments show that gas-phase reactions are not efficient enough to produce 
some terrestrial COMs such as methyl formate~\citep{Horn2004, Geppert2006, Garrod2006} in warm regions
although they may be more successful in cold regions~\citep{Balucani2015}. 
COMs such as methyl formate are believed
to be formed on warm grain surfaces by the recombination of radicals, which are products of ice mantle photodissociation
reactions,  when the temperatures of dust grains 
are between 20 and 40 K~\citep{Garrod2006}. Thus, more efficient production of COMs must rely 
on better surface astrochemical models. Progress has been made
to solve the problem of COMs formation.
In particular, the production of COMs in the three-phase model of \citet{Garrod2013} is much more efficient 
than that in the two-phase models.
However the exponential decay of photodissociation reactions with depth into ice mantles is not considered, thus
COM formation via radical-radical recombination may be overestimated.  
In a four-layer model~\citep{Kalvans2017}, the whole ice mantle is divided into 4 layers
and the shielding of species buried in ice mantle by outer layers 
is introduced. But the four-layer model is only a crude approximation; photodissociation reaction rates 
in the ice mantles should decay monolayer by monolayer. 
By modeling the layered structure of ice, we can also distinguish 
the trapped CO molecules and CO molecules that can diffuse inside ice mantles, which could provide a better answer
to the second question. \citet{Fayolle2011} suggested an extended three-phase model in which 
a mantle-surface diffusion term is introduced and not all mantle species can participate the 
mantle-surface circulation. Thus, trapped species and species that can diffuse inside ice mantles are distinguished.
However, to the best of our knowledge, 
the extended three-phase model has only been used to fit experiments~\citep{Fayolle2011}.

A numerical method to simulate surface chemistry is another issue.
It has been found that if the average population of reactive species on dust grains is much less than one, significant error may occur
if the rate equation approach is used for numerical simulation~\citep{Biham2001}, a problem known as finite size effect. 
This problem has been extensively studied in the past 20 years~\citep{Cuppen2013} and a few numerical 
methods have been suggested to replace the rate equation approach in order to solve it. The modified rate equation
method is the most efficient method, but, it also is the least rigorous approach~\citep{Garrod2008}. To the best
of our knowledge, only the modified rate equation has been adopted for hydrodynamic-chemical simulations~\citep{Furuya2015} while a
more rigorous approach, the macroscopic Monte Carlo (MC) mechanism or Gillespie algorithm~~\citep{Gillespie1976, Vasyunin2009}, has only been 
adopted to simulate the chemical evolution of simple 0-D 
collapsing cores~\citep{Vasyunin2013}.

Recently, \citet{Chang2014} reported a full gas-grain reaction network simulation with an updated surface astrochemical model,
which distinguishes the trapped species and species that can diffuse inside ice mantles.
The simulation was performed by the microscopic MC method~\citep{Chang2005}, which follows the motion of
every molecule on a grain.
The species trapped in ice mantles are called normal species while the species 
that can diffuse inside ice mantles are known as interstitial species. \citet{Chang2014} found
that many radicals are formed and then frozen inside mantles under physical conditions that pertain to cold cores.
The radicals that accumulate in the cold core stage should then be able to recombine to form COMs when the temperatures
of cores increase. It is particularly interesting to determine if COMs formed by this mechanism are able 
to explain the observed abundances of COMs toward protostars. Unfortunately, 
while microscopic MC is more rigorous 
than the macroscopic MC, its computational cost is very 
expensive when the temperatures of dust grains are higher than 15 K.
Thus, this more rigorous approach can only be used for the chemical simulations of cold cores~\citep{Chang2014}.  

In this paper, we present a macroscopic MC chemical simulation with a 1-D RHD model. 
 Our major purpose is to go one step further than earlier approaches to explain 
the COM formation during the collapse of molecular cloud cores. 
We also pay attention to the CO sublimation problem. 
We use a surface astrochemical model that is
similar to that in \citet{Chang2014} except that we use the 
macroscopic MC approach in order to reduce the computational cost. One strength of our approach 
is that as in the microscopic approach, the trapped species are also well 
distinguished from species that can diffuse inside ice mantles. 
As in \citet{Chang2014}, abundant radicals can also be formed inside the ice mantle when dust grains are about 10 K 
in our updated model. The physical model used is from \citet{Masunaga1998} and \citet{Masunaga2000}, which
has been used for simple two-phase hydrochemical simulations~\citep{Aikawa2008, Aikawa2012}.  

The remainder of our paper is organized as follows. The physical model, numerical methods and chemical model are discussed 
in Section \ref{models} while we present the simulation results of our models in Section \ref{res}.
We compare model results with observations in Section \ref{comp}. 
Discussions and conclusions follow in Section \ref{diss}.

\section{Models}\label{models}

\subsection{Physical Model}

\citet{Masunaga1998} and \citet{Masunaga2000} constructed a 1-D spherical RHD model for low mass star formation.
\citet{Aikawa2008} and \citet{Aikawa2012} calculated the molecular evolution in infalling fluid parcels in this 
1-D RHD model, which we adopt in the present work.
We refer to \citet{Aikawa2008} , \citet{Aikawa2012}, \citet{Masunaga1998} and \citet{Masunaga2000} for details of the physical 
models. We only briefly explain them below.

Initially, the number density of H nuclei at the center of a prestellar core is
$\sim6 \times 10^{4}$  cm$^{-3}$. The prestellar core is assumed to be
in hydrostatic equilibrium for a period of $10^6$ yr. We assume the temperature of the 
prestellar core before collapsing is 10 K. 
Typical lifetime of prestellar cores and triggering of the core contraction are still open 
questions in star formation studies. Decay of turbulence is one 
possibility. Then the typical lifetime of the core before the collapse 
is sound crossing time $\sim~10^6$ yr (a few 0.1 pc divided by $10^4$ cm/s). 
Alternatively, the timescale is set by the rate of mass accretion from 
the diffuse warm gas to form the filamentary molecular clouds;  
recent observational studies found that the star forming cores exist in 
filaments which have the column density higher than a critical value~\citep{Andre2014}. 
Theoretical calculations suggest that the timescale of cloud (filament) 
formation is $10^7$ yr, but it includes low (column) density phase which 
is much longer than the dense phase~\citep{Inutsuka2015}. Since 
the original hydrodynamics model we adopted~\citep{Masunaga2000}
investigates only the collapse phase, we made a simple assumption that 
the core was supported against gravitational collapse by turbulence 
and/or magnetic fields for 1 Myr before the collapse starts. 
So, at
the time $t_1 = 10^6$ yr, the prestellar core starts a 
contraction to form the first core and then a protostar.
The  protostar is formed at the time $t_2 = 10^6 + 2.5 \times 10^{5} $ yr while 
the first core is formed at around time $t_3 = t_2- 5.6\times 10^2$ yr.
After the protostar is formed, the
physical model continues to evolve for another $9.3 \times 10^{4}$ yr, thus the final time is
t$_{\textrm{final}}$ = $ 10^6 + 2.5 \times 10^{5} + 9.3 \times 10^{4}$ yr.
The model gives the physical conditions of each falling fluid parcel as a function of time. Among these physical conditions, 
the temperatures of gas and dust grains,   
density and visual extinction ($A_v$) 
are important to study the chemistry. 
Temperatures of gas and dust are set to be equal for simplicity; 
the effect of thermal decoupling of dust and gas are discussed in Section \ref{diss}.
We studied the chemical evolution of 13 fluid parcels which initially are at $r \sim 4\times 10^4$ AU. 

\begin{figure}
\centering
\resizebox{12cm}{8cm}{\includegraphics{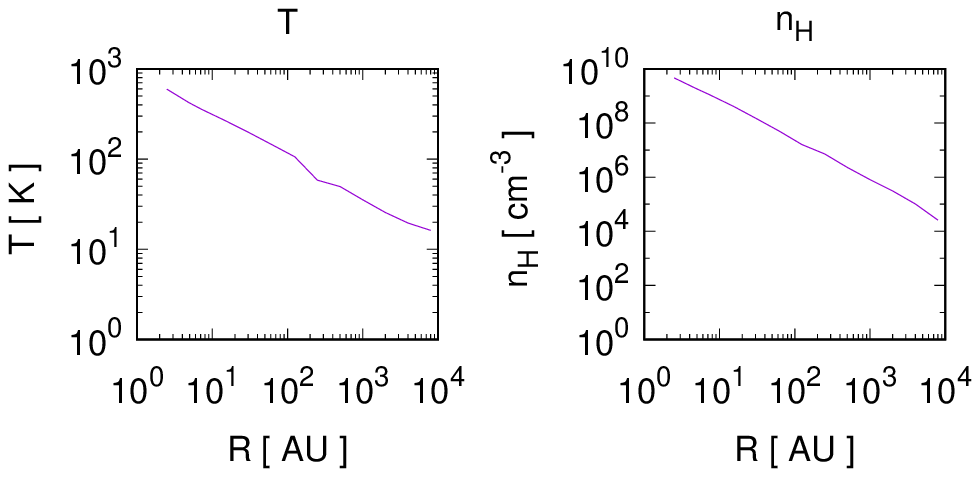}}
\caption{
Radial distribution of temperature and H nuclei density at t$_{\textrm{final}}$.
}
\label{Fig1}
\end{figure}

\begin{figure}
\centering
\resizebox{8cm}{8cm}{\includegraphics{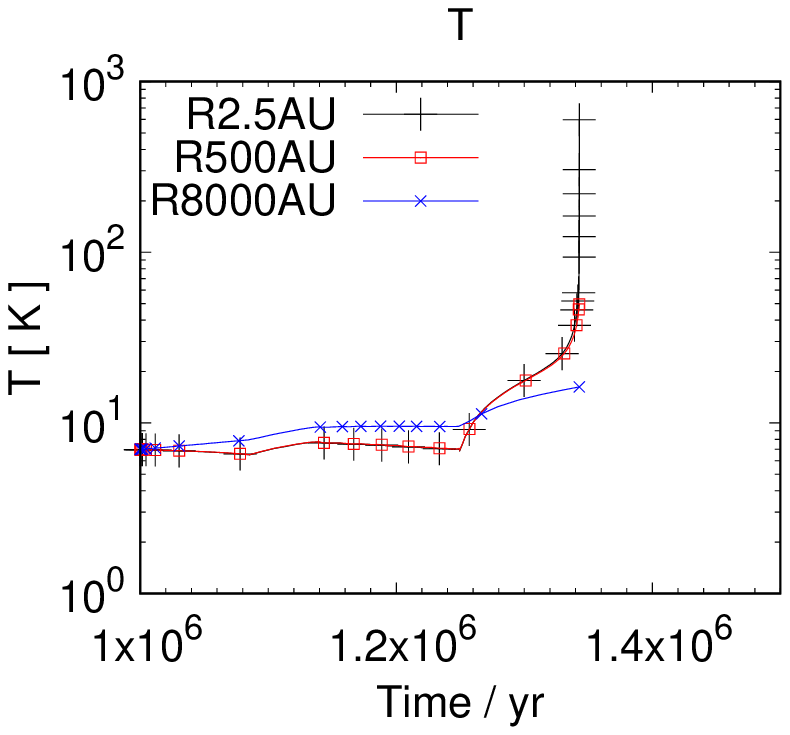}}
\resizebox{8cm}{8cm}{\includegraphics{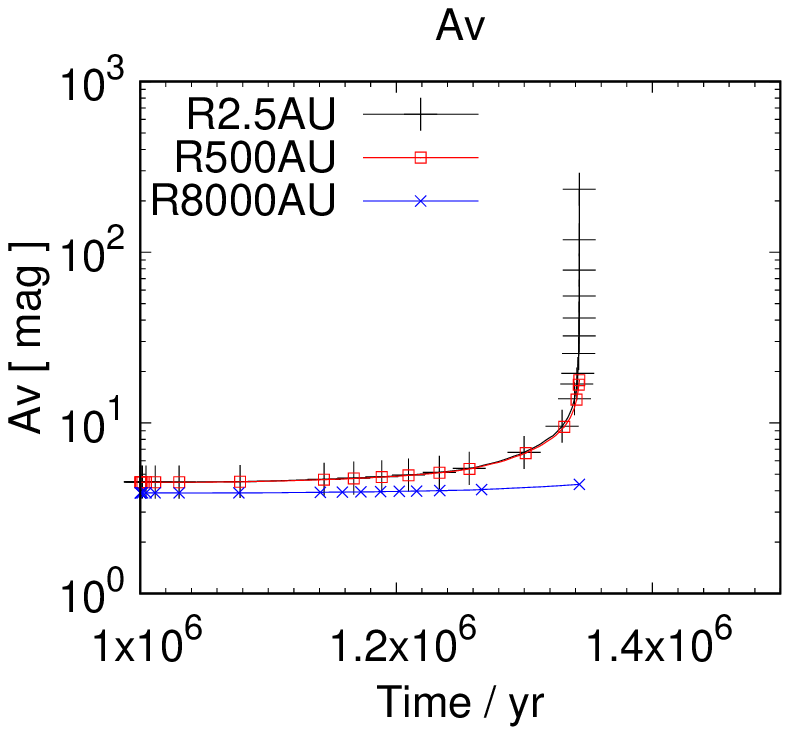}}
\resizebox{8cm}{8cm}{\includegraphics{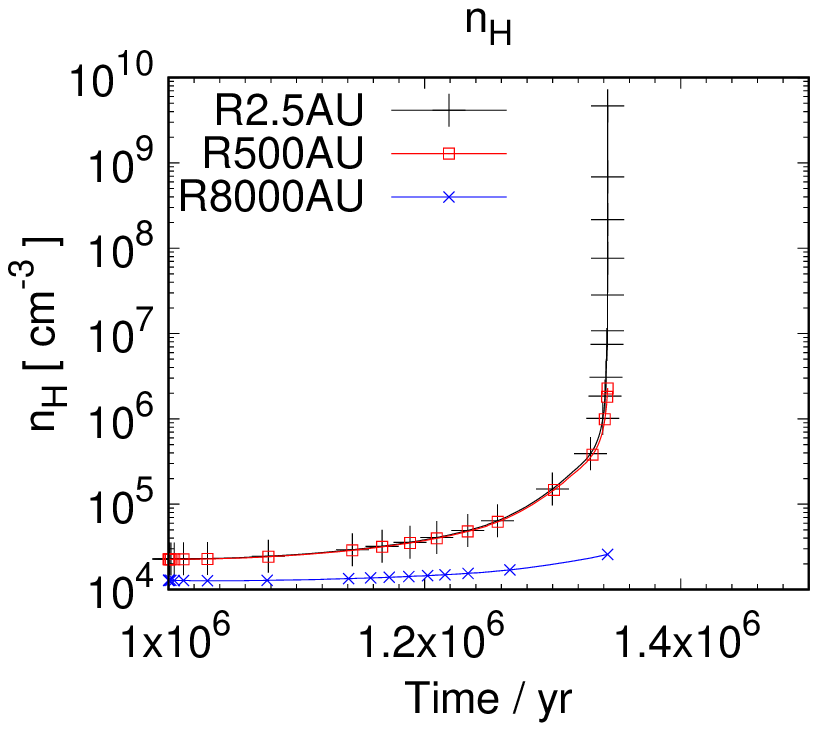}}
\caption{
The temperature, visual extinction and H nuclei density as a function of time for the fluid parcels that reach
2.5, 500 and 8000 AU at t$_{\textrm{final}}$. Physical conditions at $0<t<10^6$ yr are the same as those at $10^6$ yr.
}
\label{Fig2}
\end{figure}

Fig. \ref{Fig1} shows the radial distribution of  temperatures and H nuclei density at t$_{\textrm{final}}$. The temporal evolution of
temperature, visual extinction and H nuclei density of selected fluid parcels are shown in Fig.~\ref{Fig2}.

\subsection{Chemical Models}
The chemical reaction network used in this work is based on the network in \citet{Hincelin2011} with modifications,
which will be discussed in Section \ref{network}. 
We use the standard dust grain size  $r_d = 0.1 \mu$m where $r_d$ is the radius 
of dust grains. There are $N_s=10^6$ binding sites per monolayer on grain surfaces.
The gas-to-dust number ratio is kept at $10^{12}$. We use the standard cosmic-ray ionization rate of H$_2$, 
$\zeta=1.3 \times 10^{-17}$ s$^{-1}$. 
The surface diffusion barrier $E_b$ is fixed at 50\% of the desorption energy $E_D$.
We use the initial low-metal abundances for gas phase species~\citep{Semenov2010}, 
which are shown in Table \ref{table1}. Moreover,
we assume that initially all species are in gas phase.
Below, we first discuss our surface and bulk ice models and then move on to the reaction networks.

\begin{table}
\caption{Initial Gas Phase Abundances}
\label{table1}
\begin{tabular}{lc}
  \hline \hline
  Species & Fractional Abundance w.r.t. $n_{\rm H}$ \\
  \hline
  He       & $9.0 \times 10^{-2}$ \\
  e$^-$    & $1.2 \times 10^{-4}$ \\
  H$_2$    & $5.0 \times 10^{-1}$ \\
  C$^{+}$  & $1.2 \times 10^{-4}$\\
  N        & $7.6 \times 10^{-5}$\\
  O        & $2.6 \times 10^{-4}$\\
  S$^{+}$  & $8.0 \times 10^{-8}$\\
  Si$^{+}$ & $8.0 \times 10^{-9}$\\
  Na$^{+}$ & $2.0 \times 10^{-9}$\\
  Mg$^{+}$ & $7.0 \times 10^{-9}$\\
  Fe$^{+}$ & $3.0 \times 10^{-9}$\\
  P$^{+}$  & $2.0 \times 10^{-10}$\\
  Cl$^{+}$ & $1.0 \times 10^{-9}$\\
  \hline
\end{tabular}
\end{table}

\subsubsection{Surface and Bulk Ice Models} 

We first briefly review the multiphase model introduced by \citet{Vasyunin2013}, because
our new model of ice mantle is based on the prior model.
The multiphase model allows not only the layer-by-layer growth of ice mantles 
but also their layer-by-layer desorption. In this model, only the top 4 monolayers are active; i.e.,
all surface reactions occur only in these 4 monolayers while the bulk ice in the deep layers is completely inert.
Species inside bulk ice are frozen, so not able to diffuse to react with other species. FUV photons
are not allowed to penetrate into the ice bulk in this model, so, cannot photodissociate the ice bulk species.  
The multiphase model used a macroscopic MC approach for simulations.
We modify this multiphase model so that species inside the bulk ice are 
subject to photodissociation and the photofragments diffuse inside the bulk ice mantle.

\begin{figure}
\centering
\resizebox{7cm}{6cm}{\includegraphics{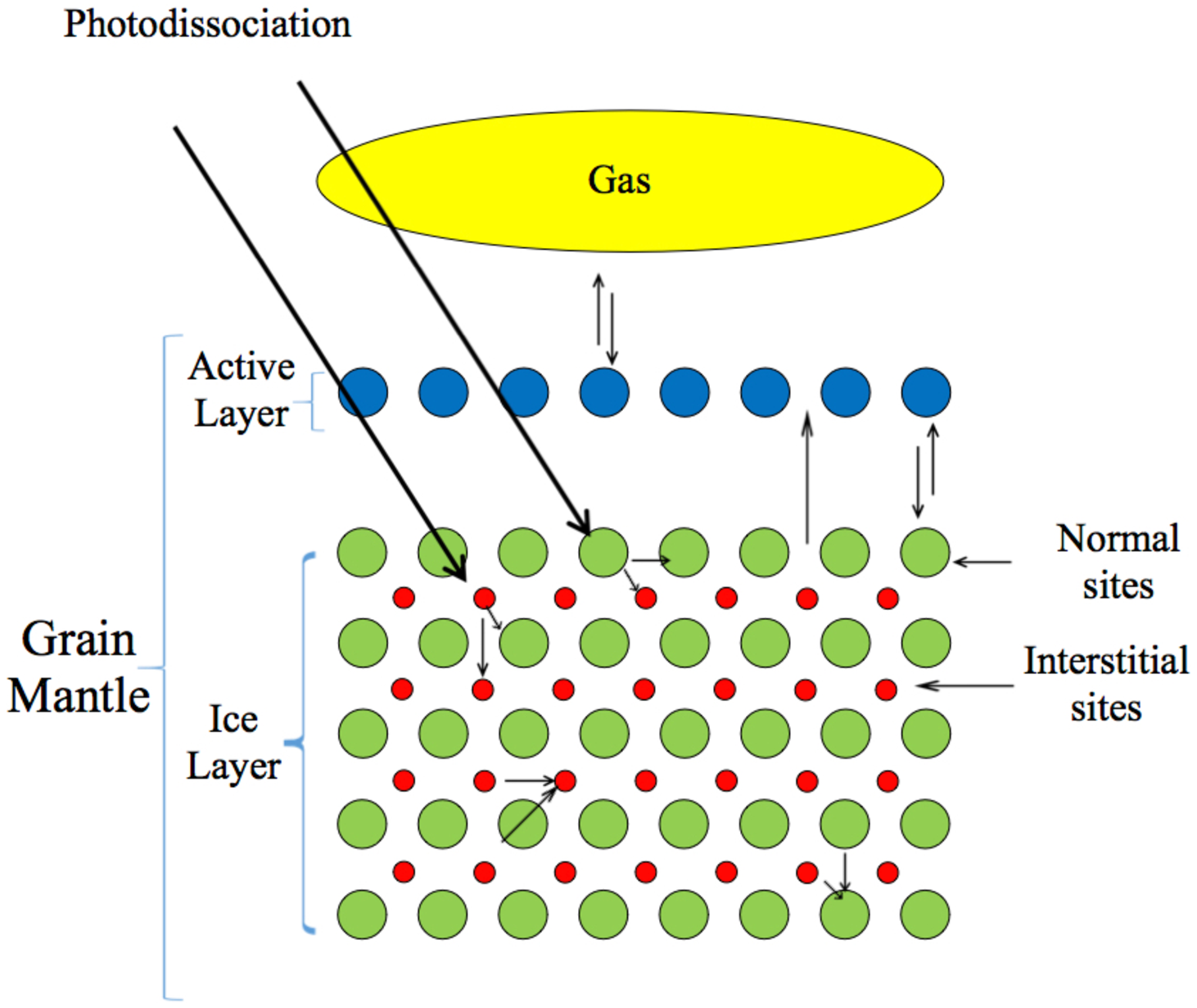}}

\caption{
Schematic diagram of the surface and bulk model.
The yellow area indicates the gas phase.
The blue small balls represent the active layer molecules, which are free to diffuse and undergo chemical reactions.
The green small balls represent the normal species in the partially active ice mantle.
They are frozen in ice mantle, thus cannot diffuse.
The red small balls represent the interstitial species. They can diffuse and react with normal or interstitial species.
Photons can penetrate into ice mantle and photodissociate species to generate interstitial species.
}
\label{Fig3}
\end{figure}

Fig. \ref{Fig3} is a schematic diagram of our surface and bulk ice model. 
We consider photon penetration, which generates photofragments in the ice mantle, and the diffusion of photofragments
in the new multiphase model.
If photons are not allowed to penetrate 
into the bulk ice mantle, our model is the same as the multiphase model of \citet{Vasyunin2013} 
since the mobile species are generated only by photons in the bulk ice in our model. 
Our model can be viewed as
a macroscopic version of the surface and bulk model in \citet{Chang2014}.
We refer the reader to \citet{Vasyunin2013} and \citet{Chang2014} for more details of the multiphase model and the microscopic
surface and bulk model.
Our macroscopic surface and bulk models have the following constraints and rules:
\begin{enumerate}
\item An ice mantle is made of surface active layers and a bulk of ice 
underneath the surface layers. The bulk ice is not inert but partly active as described below. We thus call it
partially active ice layers.

\item There are two types of binding sites, normal sites and interstitial sites,
inside the ice mantle~\citep{Chang2014}. The normal binding sites are the potential 
minima. Species that occupy normal sites are normal species. 
There are also weaker potential minima, known as subminima, which are called interstitial binding sites, 
in the partially active ice layers. Interstitial species occupy the interstitial binding sites. 
 All binding sites in the active layers are normal sites while interstitial binding sites only exist
in the bulk of ice. There are $N_s$ normal binding sites in each active layer and bulk ice layer.
The number of interstitial binding sites in each bulk ice layer is also $N_s$.

\item A bulk ice mantle is made of multiple monolayers of normal species and 
interstitial species that are uniformly distributed.
Normal species inside the bulk ice mantle are trapped (i.e. cannot diffuse) while interstitial species
are able to diffuse.

\item The active layers are made of the topmost four monolayers which can only be occupied by normal species, 
so the maximum number of normal species in the active layers is $4N_s$. When gas phase species accrete 
onto grain surfaces, they become normal species in the active layers.
The normal species in the active layers can diffuse and react with each other.

\item Surface chemical reactions and processes, including accretion, desorption and other reactions in the active layers, are 
able to change the population of the normal species in the active layers. If the total population of normal species
in the active layers is more than $4N_s$, normal species in the active layers are transformed into a normal species
in the partially active layer based on the procedure in \citet{Vasyunin2013}. Similarly, 
if the total population of normal species in the active layers is less than $4N_s$,  normal species in the partially
active ice mantle are transformed to normal species in the active layers \citep{Vasyunin2013}.
Surface photodissociation reactions in the active layers are treated in the same way as that in \citet{Vasyunin2013}.

\item We follow how photons penetrate into the partially active ice layers and photodissociate
bulk ice species in a manner similar to ~\citet{Chang2014}.
The fluxes of the external and the cosmic ray induced far-ultraviolet (FUV) photons 
are given as $F_{FUV1}=F_0 e^{-\gamma A_v}$ and $F_{FUV2}=G_0^{'} F_0$ respectively,
where $F_0=10^8$ cm$^{-2}$ s$^{-1}$ is the standard interstellar radiation field, $G_0^{'}=10^{-4}$ is the
scaling factor for cosmic ray induced photons~\citep{Shen2004} and 
$\gamma \sim 2$ measures the UV extinction relative to visual extinction \citep{Oberg2007,Roberge1991}. 
The rates of the external and the cosmic ray induced photon bombardment are
$\pi r^2 F_{FUV1}$ and $\pi r^2 F_{FUV2}$ respectively.
FUV photons penetrate into the bulk ice layer-by-layer.
At each monolayer, we randomly select a species, which is either a normal species or an interstitial 
species and then decide whether this species can be photodissociated or not. 
However, the probability to photodissociate ice species other than water by photons is poorly known.
Therefore, following \citet{Chang2014}, we calculate 
the probability that the i-th surface species is photodissociated as
$P_i = P_0 \frac{\sum k_{FUV, mi}}{\sum k_{FUV, m0}}$ where $k_{FUV, m0}$ and $k_{FUV, mi}$
are the photodissociation rate coefficients of water and the i-th surface species in the m-th product channel respectively.
Both summations are over all product channels for each species. 
The probability to photodissociate a monolayer of water was estimated to be $P_0 =0.007$~\citep{Andersson2008}.
We generate a random number $X_1$ that is uniformly distributed within (0, 1). 
If $X_1 < P_i$, the selected species is photodissociated, otherwise the FUV photon penetrates into the next monolayer.
We put the photodissociation products
in normal sites or interstitial sites based on rules 2 and 7. The exponential decay
of photodissociation rates in ice mantles is mimicked well~\citep{Chang2014}. 

\item Interstitial species are generated when FUV photons photodissociate normal species in the partially active ice mantle.
Photodissociation products of a normal species can include interstitial species, and thus, the normal sites which
the photodissociated normal species occupied can become empty. 
Assuming there are $N_{nh}$ and $N_{ih}$ empty normal and interstitial binding 
sites respectively in a monolayer, the probability
that a photodissociation product occupies a normal binding site is, 
$p_n = \frac{N_{nh}}{N_{nh}+ \alpha N_{ih}}$, where $\alpha$ is a parameter that differentiates 
the normal binding sites and interstitial binding sites. 
The smaller the $\alpha$  value is,
the more likely photodissociation products occupy normal sites.
 We assume that species occupying the normal sites are always no fewer 
than those in interstitial sites in the ice mantles
because the normal binding sites are the potential minima while the 
interstitial binding sites are the potential subminima. The value of $\alpha$ is therefore $\leq 1$.
If $\alpha = 1$, the potential minima is the same as the potential subminima, which is an extreme case. 
On the other hand, if $\alpha = 0$, photodissociation products are prohibited from occupying interstitial sites.
Photodissociation reactions increase the population of species, and therefore, there are not enough normal sites for the
extra species to occupy if $\alpha =0$. So, we limit the range of $\alpha$ to be $0< \alpha \leq 1$.  
However, $\alpha$ is poorly constrained other than the discussions above. 
We set $\alpha=0.01, 0.5$ in our models to test the influence of its value on simulation results.
In the MC simulations, a random 
number X, which is uniformly distributed within 0 and 1, is generated. If $X< p_n$, the photodissociation product 
occupies normal sites. Otherwise, it occupies interstitial sites.

\item Interstitial species are able to diffuse into the active layers from the partially active bulk of ice.
Assuming the bulk diffusion barrier is $E_{b2}$, 
the rate that an interstitial species hops from one interstitial site to another
one is $k_{hop}= \nu \exp(-\frac{E_{b2}}{T})$, where $T$ is the grain surface 
temperature and $\nu=10^{12}$ s$^{-1}$ is the trial frequency.
See below the discussion about the bulk diffusion barriers. 
Two assumptions are made in order to derive the rate at which interstitial species 
diffuse into the active layers. First, interstitial species can hop toward 6 different 
directions~\citep{Chang2014} while 
only the hopping toward the active layers transforms an interstitial species into a normal species in the active layers. 
Second, only the interstitial species that are in the monolayer which is next to the active 
layers can diffuse into the active layers. The rate at which the jth interstitial species 
diffuse into the active layers is thus, 
\begin{equation}
r_{diff_j} = \frac{k_{hop_j}}{6} \frac{N_j}{N_L},
\label{diffuse}
\end{equation}
where $N_L$ is the number of monolayers of normal species in the partially active ice mantle, 
$N_j$ is the population of the jth interstitial species and $k_{hop_j}$ is the hopping rate of the jth interstitial species.
In the active layers, the population of normal species increases when interstitial species 
diffuse into these layers. We follow the rules 
in \citet{Vasyunin2013} to move one normal species in the active layers to the partially active ice mantle 
if there are more than $4N_s$ species in the active layers.

\item Interstitial species are able to react with other interstitial species or normal species in the partially active ice mantle.  
The reaction rate between two interstitial species can be derived in a manner similar to that for surface reactions.
The rate coefficient of a reaction between two surface species is
based on the assumption that two species must on average visit all sites on grain surfaces
before they encounter each other~\citep{Hasegawa1992}. We assume that two interstitial species must also on average visit all interstitial sites in
the partially active ice mantle before they encounter each other. So, the rate of the reaction between two different
interstitial species is,  
\begin{equation}
r_{ij} = \frac{\kappa_{ij}}{N_s N_L} (k_{hop_i} + k_{hop_j})N_i N_j,
\label{int_int}
\end{equation}
where $N_L$ is the number of monolayers in the partially active ice mantle,
$N_i$  and $N_j$ are the population of the ith and jth interstitial species respectively, 
$\kappa_{ij} \leq 1$ is a factor for the reaction activation energy~\citep{Semenov2010,Chang2007,Garrod2011},
and $k_{hop_i}$ and $k_{hop_j}$ are the hopping rates of the ith and jth interstitial species respectively.
The factor $\kappa_{ij}$ is not a simple exponential function to overcome the barrier;
we consider the tunneling and competition between diffusion and reaction.
The rate of the reaction between two identical species i is
\begin{equation}
r_{ii} = \frac{\kappa_{ii}}{2N_s N_L} (k_{hop_i} + k_{hop_i})N_i^2.
\label{int_int_ii}
\end{equation}
Similarly, because all normal species in the partially active ice mantle are frozen and only interstitial species
can diffuse in the partially active ice mantle, 
the rate of the reaction between an interstitial species i and a normal species n is,
\begin{equation}
r_{in} = \frac{\kappa_{in}}{N_s N_L} k_{hop_i} N_i N_n,
\label{int_nor}
\end{equation}
where $N_n$ is the population of the nth normal species.
When interstitial species react with normal species, the product becomes normal species
if there is only one product. Otherwise, the product with the largest diffusion barrier 
becomes a normal species while all other products become interstitial species.
 
\end{enumerate}

We simulate 5 different models as shown in Table~\ref{table2}. 
Hereafter, our new surface and bulk model are called the new multiphase model.
In models MC1, MC2 and MC3, we adopt the new multiphase model. 
Normal species are frozen in partially active ice mantles while 
the bulk diffusion barrier $E_{b2}$ for interstitial species has been poorly constrained so far. Following ~\citet{Chang2014}, 
we set $E_{b2}= 0.7 E_D$ in models MC1 and MC2 while in model MC3, we set $E_{b2}= 1.0 E_D$.
The desorption energy of species in the active layers are taken from~\citet{Garrod2006}.
The parameter $\alpha$ is even less constrained. In model MC1, $\alpha$ takes a value of 0.01 assuming
photodissociation products in the partially active ice mantle can hardly occupy interstitial sites while 
in models MC2 and MC3, $\alpha$ takes a value of 0.5 so that much more photodissociation products
can occupy interstitial sites. 
In model MC4, we use the basic multiphase model proposed by \citet{Vasyunin2013} while MC5 is a two-phase model.
The basic multiphase model is similar to the three phase models proposed by~\citet{Taquet2012} 
and \citet{Furuya2015} in that the ice mantle consists of chemically active surface layers and inert ice bulk mantle
or photofragments are assumed to recombine immediately in the bulk ice mantle.
In model MC5, all surface species are in the active layers and 
there is no inert or partially active ice bulk.
The photodissociation rates of surface species may be modified so that
the exponential decay of photodissociation rates with depth into ice mantle is included in two-phase models.
This more rigorous treatment of surface photodissociation reactions
is not considered in most two-phase astrochemical models. Therefore, we do not take into account
the exponential decay of photodissociation reaction rates in 
two-phase models in this paper for simplicity.
Models MC4 and MC5 are simulated for the purpose of comparison with models MC1, MC2, MC3.

\begin{table}
\caption{Parameters for Different Models}
\label{table2}
\begin{tabular}{cccc}
  \hline \hline
  Model & $E_{b2}$/$E_D$ & $\alpha$ & Surface and Bulk Model \\
  \hline
  MC1   &   0.7 & 0.01 & new multiphase \\
  MC2   &   0.7 & 0.5 & new multiphase \\
  MC3   &   1   & 0.5 & new multiphase\\
  MC4   &   -   & - & basic multiphase\\
  MC5   &   -   & - & two-phase \\
  \hline
\end{tabular}
\end{table}

\subsubsection{Reaction Network}\label{network}

Radicals CH$_3$O and CH$_2$OH are not distinguished in the original reaction network~\citep{Hincelin2011}.
A major modification is to distinguish these two radicals.

Photodissociation reactions are the major formation pathways
of CH$_2$OH and CH$_3$O~\citep{Chang2016}. 
In the active layers or the partially active bulk ice mantle, both CH$_3$O and CH$_2$OH can be formed
by photodissociation of methanol while CH$_2$OH can also be produced by 
the hydrogenation of H$_2$CO. Photodissociation of methanol can also produce CH$_3$.
The methanol photodissociation branching ratios to produce CH$_3$O, CH$_2$OH and CH$_3$
are 20:20:60~\citep{Garrod2008}. 
Although the hydrogenation of both surface or bulk CH$_3$O and CH$_2$OH forms methanol,   
they can react with other radicals to form different COMs as in the following reactions:
\begin{equation}
\begin{array}{lllll}
\rm{CH}_3\rm{O} & + & \rm{HCO} & \rightarrow & \rm{HCOOCH}_3 \\
\rm{CH}_3\rm{O} & + & \rm{CH}_3 & \rightarrow & \rm{CH}_3\rm{OCH}_3 \\
\rm{CH}_2\rm{OH} & + & \rm{HCO} & \rightarrow & \rm{CH}_2\rm{OHCHO} \\
\rm{CH}_2\rm{OH} & + & \rm{CH}_3 & \rightarrow & \rm{C}_2\rm{H}_5\rm{OH}. \\
\end{array}
\end{equation}
Both glycolaldehyde (CH$_2$OHCHO) and methoxy (CH$_3$O) are new species in the reaction network. Relevant glycolaldehyde
reactions are included in the reaction network~\citep{Garrod2015}. We assume that gas phase CH$_3$O can participate
in all reactions in which gas phase CH$_2$OH participates. The rate coefficients of the
reactions between gas phase CH$_3$O or CH$_2$OH and other species are assumed to be the same. 
The binding energies of methoxy and glycolaldehyde are not available in~\citet{Garrod2006},
but are set to be 5084 K and 6295 K respectively because we assume the binding energies 
of isomers are the same in our models. 
Table~\ref{table22} shows the desorption energies, $E_D$ of selected surface species.

The encounter desorption mechanism is used in this work~\citep{Hincelin2015}. We include the reaction
gH$_2$ + gH$_2$ $\rightarrow$ gH$_2$ + H$_2$ in the active layers where the letter g designate
species in these layers. As the density of molecular clouds increases, too many gH$_2$ molecules are
frozen on grain surfaces without the encounter desorption mechanism, which is not physical. 
The encounter desorption mechanism takes into account the fact that the desorption 
energy of gH$_2$ on another gH$_2$ molecule
is much smaller than that of gH$_2$ on water ice,  which significantly increases the desorption of gH$_2$.  
\begin{table}
\caption{The Desorption Energies of Selected Species}
\label{table22}
\begin{tabular}{cc}
  \hline \hline
  Species & $E_D$(K) \\
  \hline
  H        &  450 \\ 
  O        &  800\\
  CO       &  1150 \\
  CO$_2$   &  2575 \\
  N$_2$    &  1000 \\
  HCO      &  1600 \\
  H$_2$CO  &  2050 \\
  CH$_3$   &  1175 \\
  CH$_2$OH &  5084 \\ 
  CH$_3$O  &  5084 \\
  HCOOCH$_3$ & 6295 \\
  CH$_2$OHCHO & 6295 \\
  \hline
\end{tabular}
\tablecomments{Taken from~\citet{Garrod2006}. The desorption energies of isomers are assumed to be the same.}
\end{table}

Finally, the CO hydrogenation reactions on grain surface to form methanol have been well studied in laboratory experiments.
The microscopic MC method has been used to fit the activation barriers for the reactions CO + H $\rightarrow$ HCO 
and H$_2$CO + H $\rightarrow$ H$_3$CO~\citep{Fuchs2009}, in which the competition 
mechanism~\citep{Chang2007,Chang2012,Garrod2011} can be naturally included. The 
competition mechanism and the reaction barriers 
fitted by \citet{Fuchs2009} are used to calculate the rates for surface reactions H$_2$CO + H 
and CO + H in this work. 
However, the reaction rates are only fitted at fixed temperatures, 
T = 12 K, 13.5 K, 15.0 K and 16.5 K~\citep{Fuchs2009} respectively.
In order to perform astrochemical simulations, we assume that the reaction rates
are constant at temperature ranges, 10.0 K $\leq$ T $\leq$ 12 K, 12 K $<$ T $\leq$ 13.5 K,
13.5 K $<$ T  $\leq$ 15.0 K and T $\geq$ 15 K.
For T $\leq$10 K, the factor $\kappa_{ij}$ is assumed to be constant.
For all other reactions with barriers on grain surfaces, the rate coefficients 
are calculated based on tunneling as in the original reaction network~\citep{Hincelin2011}. 

In total, there are 663 species and 6370 chemical reactions in our chemical models
; among them, 464 species and 4667 reactions are in the gas phase.

\subsection{Numerical Methods}
An accelerated Gillespie algorithm (QSSA1), which is based on the regular Gillespie algorithm, 
is used in this work~\citep{Chang2017}. We first briefly introduce the regular Gillespie algorithm,  
and then the accelerated Gillespie algorithm method.

In order to perform simulations with the Gillespie algorithm,  
we isolate a cell of gas containing one dust grain~\citep{Vasyunin2009}.
There are $10^{12}$ H nuclei in the cell of gas because we assume the gas-to-dust number ratio is $10^{12}$. 
The initial population of species a is $fr_a 10^{12}$ where
$fr_a$ is the initial fractional abundance of species a.
The reaction rates of all 
chemical reactions in gas phase and on dust grains
must be calculated. For the m-th bimolecular gas-phase or surface reaction with two different species a and b, 
the reaction rate is, $r_m = k_m N_a N_b$, where $k_m$ is the rate coefficient of the m-th
reaction, $N_a$ is the population of species a and $N_b$ is the population of species b. 
If the reactants in the n-th bimolecular reaction are the identical species a, 
the reaction rate is $r_n = 0.5 k_n N_a(N_a-1)$.
For the m-th unimolecular reaction with reactant a, the reaction rate is $r_m =  k_m N_a $.

The numerical implementation of the regular Gillespie algorithm is the following. First, the total reaction rate of
all reactions is calculated by $r_{\textrm{total}} = \sum_m r_m$ where m is for all reactions in the chemical system.
The time step is calculated by $\Delta t = -ln(X)/r_{total}$ where $X$ is a random variable uniformly distributed
in the range between 0 and 1. Assuming the current time is $t_0$, a reaction will fire at the time $t_{\textrm{next}} = t_0 + \Delta t$.
The n-th reaction, which fires at $t_{\textrm{next}}$, satisfies the equation
$\sum_{m=0}^{m=n-1} r_m < r_{\textrm{total}}Y \leq \sum_{m=0}^{m=n} r_m$ where
$Y$ is another random variable uniformly distributed between 0 and 1. The populations of all reactants
involved in the n-th reaction are updated at the time $t_{\textrm{next}}$. 
Immediately after the n-th reaction fires,
the reaction rates of all reactions in which the population of reactants changes are also updated 
at the time $t_{\textrm{next}}$. The processes are repeated
until $t_{\textrm{next}}$ reaches the final time.     
 
We used the accelerated Gillespie algorithm instead of the regular Gillespie algorithm to perform simulations 
in this work because H$_2$ accretion and desorption consume significant
amounts of CPU time if the regular Gillespie algorithm is used. The accelerated Gillespie algorithm is based
on quasi-steady-state assumption~\citep{Rao2003}.  
The implementation of this Gillespie algorithm is similar to that of the regular Gillespie algorithm. 
The difference is that the conditional 
expectation of reaction rates are used to calculate the time step $\Delta t$ and the selection of the 
n-th reaction which fires at $t_{\textrm{next}}$.
We treat surface H$_2$ as a transient                                                  
species in the accelerated Gillespie algorithm~\citep{Chang2017}.
The conditional reaction rates of all reactions that do not involve surface and gas phase H$_2$ are the
same as the reaction rates in the regular Gillespie algorithm. However, the conditional rates of reactions
that involve gas phase and surface H$_2$ are calculated using a different algebraic expression, in which
the accretion of gas phase H$_2$, thermal desorption and encounter desorption of surface H$_2$ do not 
explicitly appear in the gas-grain reaction network, so that the simulations are significantly accelerated.   
We refer to \citet{Chang2017} for details of the accelerated Gillespie algorithm.
Finally, when the temperature of the fluid parcel exceeds 200 K, 
the molecular evolution of fluid parcels is dominated by gas phase chemistry.
We thus stop the accretion of all species in order to
further increase the simulation efficiency. 

We use the Taurus High Performance Computing system of Xinjiang Astronomical Observatory for
the simulations in this work. The computational cost varies depending on the fluid parcel and the chemical model.
It takes no more than 3 days to simulate the molecular evolution of each fluid parcel for each chemical model.

\section{Results}\label{res}

\subsection{Evolution of Granular and Gas-Phase Species in Fluid Parcels}

\begin{figure}
\centering
\resizebox{12cm}{12cm}{\includegraphics{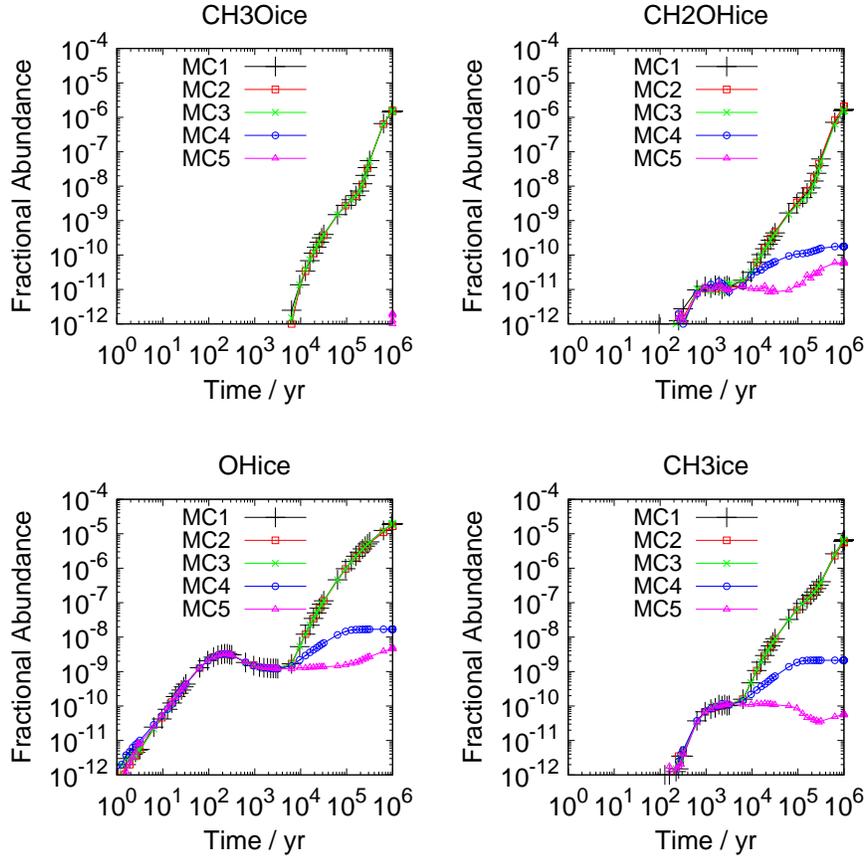}}
\caption{
Selected granular radical abundances as a function of time in the fluid parcel that reaches r=2.5 AU at $t_{\textrm{final}}$
before the contraction of the molecular cloud in models MC1, MC2, MC3, MC4 and MC5. The temperature is 
fixed at 10 K.
}
\label{Fig4}
\end{figure}

Fig.~\ref{Fig4} shows the temporal variation of selected radicals JCH$_3$O, JCH$_3$, JCH$_2$OH and JOH 
before the contraction of the molecular cloud in the fluid parcel 
that reaches r= 2.5 AU at the final time in models MC1, MC2, MC3, MC4 and MC5. 
The letter J designates granular species, which are the species on dust grains. In models MC1, MC2 and MC3, 
granular species Ji includes species gi,
interstitial species i (Ii) and species i locked in the normal sites (Ki). 
In model MC4, granular species Ji includes species gi  
and species i in the inert ice bulk while in model MC5 granular species Ji is equivalent to species gi.
We can see that the abundances 
of these radicals are almost the same before around $7\times 10^3$ yr in all models. This is because there are fewer than 4 
monolayers of granular species on dust grains before about this time, so all granular molecules in models 
MC1, MC2, MC3 and MC4 are in the active layers, so that surface chemistry in models MC1, MC2 and MC3 and MC4 is the same 
as that in MC5. After $7\times 10^3$ yr, the differences of granular radical abundances among
models increase dramatically.
In models MC1, MC2 and MC3, radicals can be generated by photodissociation reactions or recombination reactions  
and then buried in the partially 
active ice bulk.
Interstitial species can also diffuse to enter the active layers. Among them, IH atoms have the highest diffusion rate.  
Radicals that diffuse much slower than IH or these that are frozen in the normal sites tend to accumulate in the ice mantle.
On the other hand, some recombination reactions in the bulk of ice can also produce radicals. For instance, 
IH can recombine with IO or KO to form IOH or KOH.
So, granular radical abundances 
increase quickly after $7\times 10^3$ yr.
The rapid increase of granular radical abundances in models MC1, MC2 and MC3 is similar to that in 
the microscopic Monte Carlo models~\citep{Chang2014}. 
The differences in granular radical abundances among the new multiphase models MC1, MC2 and MC3 are not significant,
because radicals are mainly formed via photodissociation of ice molecules.
Photofragments of major icy species such as JO, JOH and JCH$_3$ are the dominant granular radicals.
The population of all radicals account for 20-30\% of the total population of granular species in models MC1-3. 
In the two-phase model MC5, on the other hand, radicals produced on grain surfaces
quickly react with newly accreted species such as JH; thus, granular radicals cannot accumulate in model MC5.
At the time of $10^6$ yr, the abundances of granular radicals in models MC1, MC2 and MC3 
are more than four orders of magnitude higher than those in model MC5. 
The abundances of granular radicals in the basic multiphase model MC4 are typically higher than those in model MC5 
but much lower than those in our multiphase models. At $10^6$ yr, the abundance of JCH$_3$O in model MC4 
is more than five orders of magnitude lower than those in our new multiphase models.
The abundances of JCH$_3$, JCH$_2$OH and JOH in model MC4 are more than a factor of 4 higher than those in model MC5 but
a few orders of magnitude less than that in our new multiphase models. The larger granular radical 
abundances in model MC4 than model MC5 occurs because surface radicals can be buried in the inert ice mantle if these radicals 
do not react with other species in the active layers.

\begin{figure}
\centering
\resizebox{12cm}{12cm}{\includegraphics{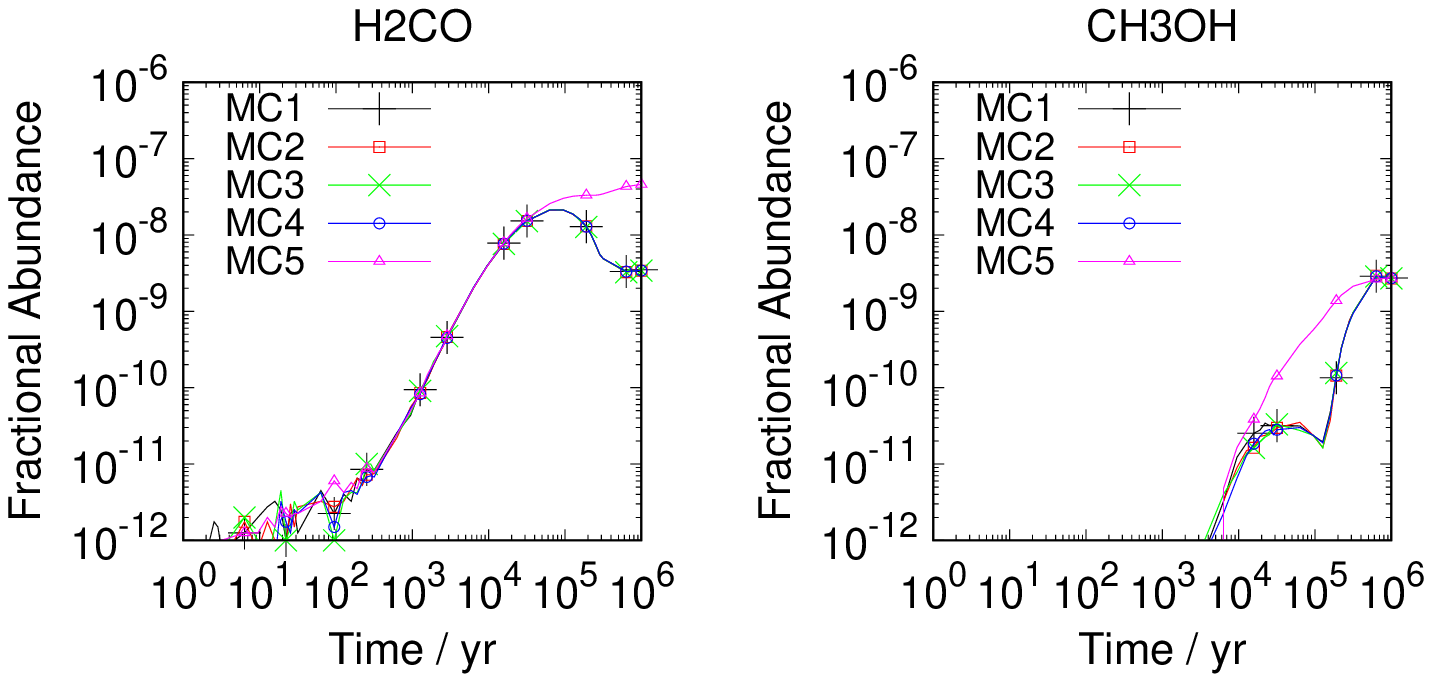}}
\caption{
Selected gas-phase species abundances as a function of time in the fluid parcel that reaches r=2.5 AU at $t_{\textrm{final}}$
before the contraction of the molecular cloud in models MC1, MC2, MC3, MC4 and MC5. The temperature is 
fixed at 10 K.
}
\label{Fig5}
\end{figure}

Fig.~\ref{Fig5} shows the temporal evolution of selected gas-phase species before the start of contraction 
in the fluid parcel
that reaches r= 2.5 AU at $t_{\textrm{final}}$. 
The visual extinction of the fluid parcel is around 4 before the start of contraction, 
thus the FUV radiation is intense enough for
the photodesorption to play important roles in the production of gas-phase H$_2$CO and CH$_3$OH in our models.  
On the other hand, thermal desorption is not important before the start of contraction because the temperature is fixed at 10 K.
We can see that the abundances of gaseous H$_2$CO and CH$_3$OH predicted by the basic and new multiphase models 
are similar because photons can only desorb species in the active layers, and because the population of gH$_2$CO and gCH$_3$OH  
are similar in these models. Before $2\times 10^4$ yr, the abundances of gaseous H$_2$CO and CH$_3$OH predicted
by the two-phase model are also similar to those by models MC1-4 because all granular species are in the active layers in
models MC1-4. After $2\times 10^4$ yr, the two-phase model predicts higher H$_2$CO abundance because all
JH$_2$CO can be photodesorbed in the two-phase model while only gH$_2$CO can be desorbed in models MC1-4.   
The two-phase model also predicts higher CH$_3$OH abundance than other models do between $2\times 10^4$ and $5 \times 10^5$ yr
because the abundance of JCH$_3$OH in the two-phase model is higher than gCH$_3$OH abundances in models MC1-4.
However, after $5 \times 10^5$ yr all models predict similar CH$_3$OH abundances because the abundance of JCH$_3$OH in the 
two-phase model and gCH$_3$OH abundances in models MC1-4 are similar.

\begin{figure}
\centering
\resizebox{12cm}{12cm}{\includegraphics{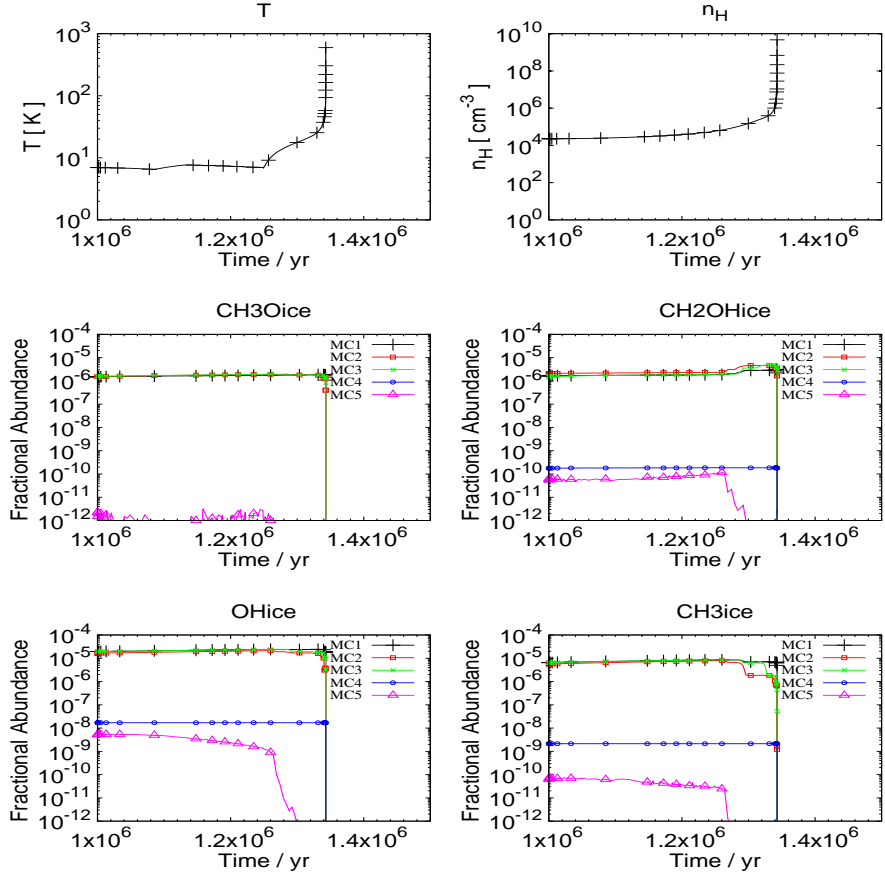}}
\caption{
The temporal variation of selected granular radical abundances 
 after the molecular clouds start to contract
in the fluid parcel that reaches r=2.5 AU at $t_{\textrm{final}}$ 
in models MC1, MC2, MC3, MC4 and MC5.
The temporal evolution of the temperature and density for the fluid parcel is also plotted.
}
\label{Fig6}
\end{figure}

Fig.~\ref{Fig6} shows the temporal evolution from the start of contraction of abundances for 
the above selected granular radicals in the fluid parcel that reaches r=2.5 AU at $t_{\textrm{final}}$ in all models. 
The temporal evolution of the temperature and density for the fluid parcel is also plotted as well.
During the contraction, the temperature of the fluid parcel first slightly decreases and then quickly increases after around $1.26\times 10^6$ yr.
We can see that granular radical abundances as a function of time vary among models. 
In the two-phase model MC5, granular radicals quickly disappear after $1.26\times 10^6$ yr because these
granular radicals are able to diffuse to react with other radicals or desorb into gas phase.  
Almost all granular radicals in model MC4 are buried 
in the completely inert ice mantle before most water molecules desorb at $\sim 1.34\times 10^6$ yr,
so the granular radical abundances do not change much before this time. 
After most water molecules desorb, the ice mantle no longer exists and 
granular radicals quickly disappear in model MC4. 
The temporal evolution of granular
radicals in the new multiphase models MC1, MC2 and MC3 is more complicated. 
Before $1.26\times 10^6$ yr, when the temperature of the fluid parcel is $\lesssim$ 10 K,
radicals can still be produced in the partially active ice mantle, so the abundances of granular radicals slightly increase. 
When the temperatures quickly increase after $ 1.26\times 10^6$ yr, granular radicals are able to diffuse inside
the partially active ice mantle and more interstitial species can react with other species or diffuse into the active layers. 
The recombination of granular radicals obviously can reduce the abundances of these radicals, but may also produce
other granular radicals.
We can see that the JCH$_2$OH abundance first increases after $1.26\times 10^6$ yr and then remains almost constant because
JCH$_2$OH  can be produced by the recombination of species in the ice mantle by reactions such as
JCH$_3$ + JO $\rightarrow$ JCH$_2$OH, which is barrierless in our reaction network. The production rate exceeds
the destruction rate for JCH$_2$OH inside the ice mantle, so the abundance of JCH$_2$OH initially increases
with temperature in models MC1, MC2 and MC3.      
However, after $1.34\times 10^6$ yr, when most water molecules desorb from grain surfaces, JCH$_2$OH quickly disappears
in our new multiphase models as in the basic multiphase model MC5. 
The production rates of JCH$_3$O, JCH$_3$ and JOH are 
less than the loss rates of these radicals inside ice mantles, so
their abundances decrease after $1.26\times 10^6$ yr in our new multiphase models.

\begin{figure}
\centering
\resizebox{12cm}{12cm}{\includegraphics{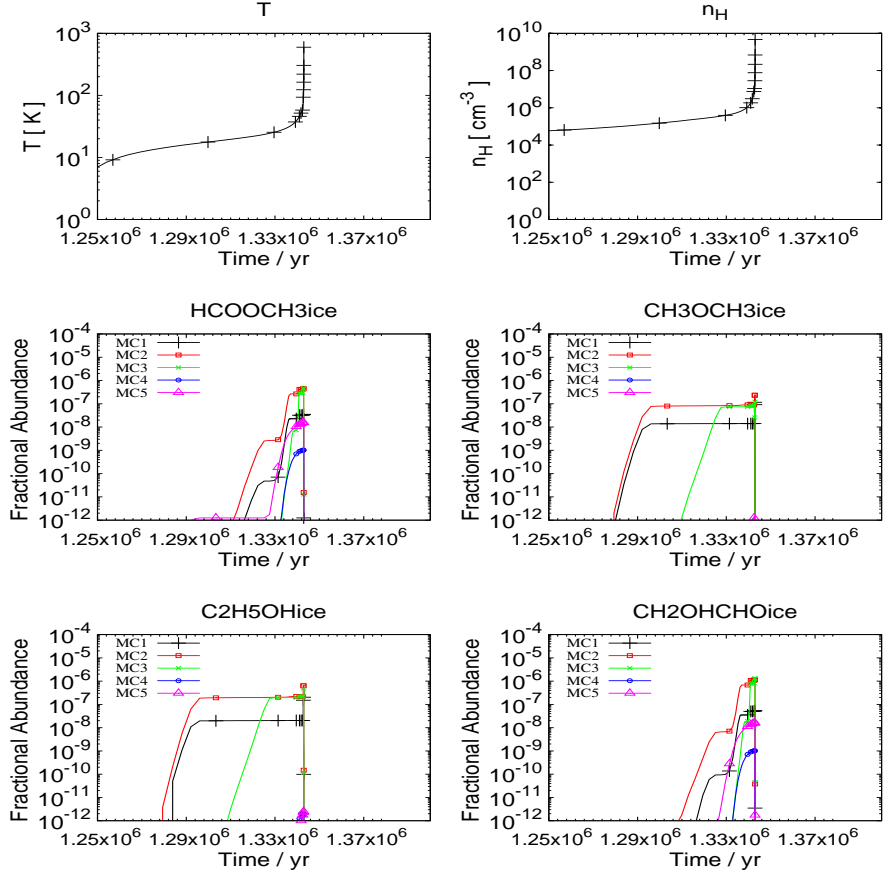}}
\caption{
The temporal variation of selected COM abundances in the fluid parcel that reaches r=2.5 AU at $t_{\textrm{final}}$ 
after the temperatures of the fluid parcel starts to increase in models MC1, MC2, MC3, MC4 and MC5.
The temporal evolution of the temperature and density for the fluid parcel is also plotted.
}
\label{Fig7}
\end{figure}

Significant amounts of COMs are produced by granular radicals recombination inside the 
partially active ice mantle in models MC1, MC2 and MC3 as shown in Fig.~\ref{Fig7}.
As in Figure \ref{Fig6}, the temporal evolution of the temperature and density for the 
fluid parcel is also plotted in Fig.~\ref{Fig7}.
We can see that the abundance of JHCOOCH$_3$ in models MC2 and MC3 is almost two orders of magnitude higher than that in 
model MC5 around the time of $1.33\times 10^6$ yr. Methyl formate molecules are formed by the recombination of the 
two radicals JHCO and JCH$_3$O. In the two-phase model MC5, radical abundances are much lower than in our 
new multiphase models, so that the production rate of COMs is much lower than that 
in models MC2 and MC3. We can also see that the efficient production
of JHCOOCH$_3$ occurs earlier in model MC2 than in model MC3 because the bulk diffusion barriers in model
MC2 are lower, so interstitial radicals start to diffuse at lower temperatures before the temperatures of the
fluid parcel further increases in model MC2. On the other hand, we can see that the $\alpha$ value 
can also significantly influence the production of granular COMs. The abundance of JHCOOCH$_3$ in model MC1 
is about one order of magnitude lower than in models MC2 and MC3. Because $\alpha$ is smaller in model MC1, 
radicals produced by photodissociation reactions are less likely to occupy interstitial sites 
than in models MC1 and MC2. Since only interstitial species
can diffuse to react with other species in our new multiphase models, 
the production of COMs in model MC1 is less efficient.
The binding sites where granular COMs are formed also varies among models.
Because the products of the reactions by two interstitial radicals still occupy interstitial sites in our new multiphase models,
there are more interstitial COMs in models MC2 and MC3 than in model MC1. 
The ratio of the population of interstitial JHCOOCH$_3$
to the population of normal JHCOOCH$_3$ is about 1:4, 2:1 and 2:1 in models MC1, MC2 and MC3 respectively at 
$1.34\times 10^6$ yr, when the temperature is about 44 K. 
Formation of JHCOOCH$_3$ in model MC4 is less efficient than that in other models, because only radicals in the active layers 
can diffuse to recombine and form JHCOOCH$_3$ while buried radicals in the inert ice mantle cannot diffuse. 
Similarly, we can explain the difference in the production of JCH$_3$OCH$_3$, JC$_2$H$_5$OH and JCH$_2$OHCHO 
inside the ice bulk mantle among the models. 

\begin{figure}
\centering
\resizebox{12cm}{12cm}{\includegraphics{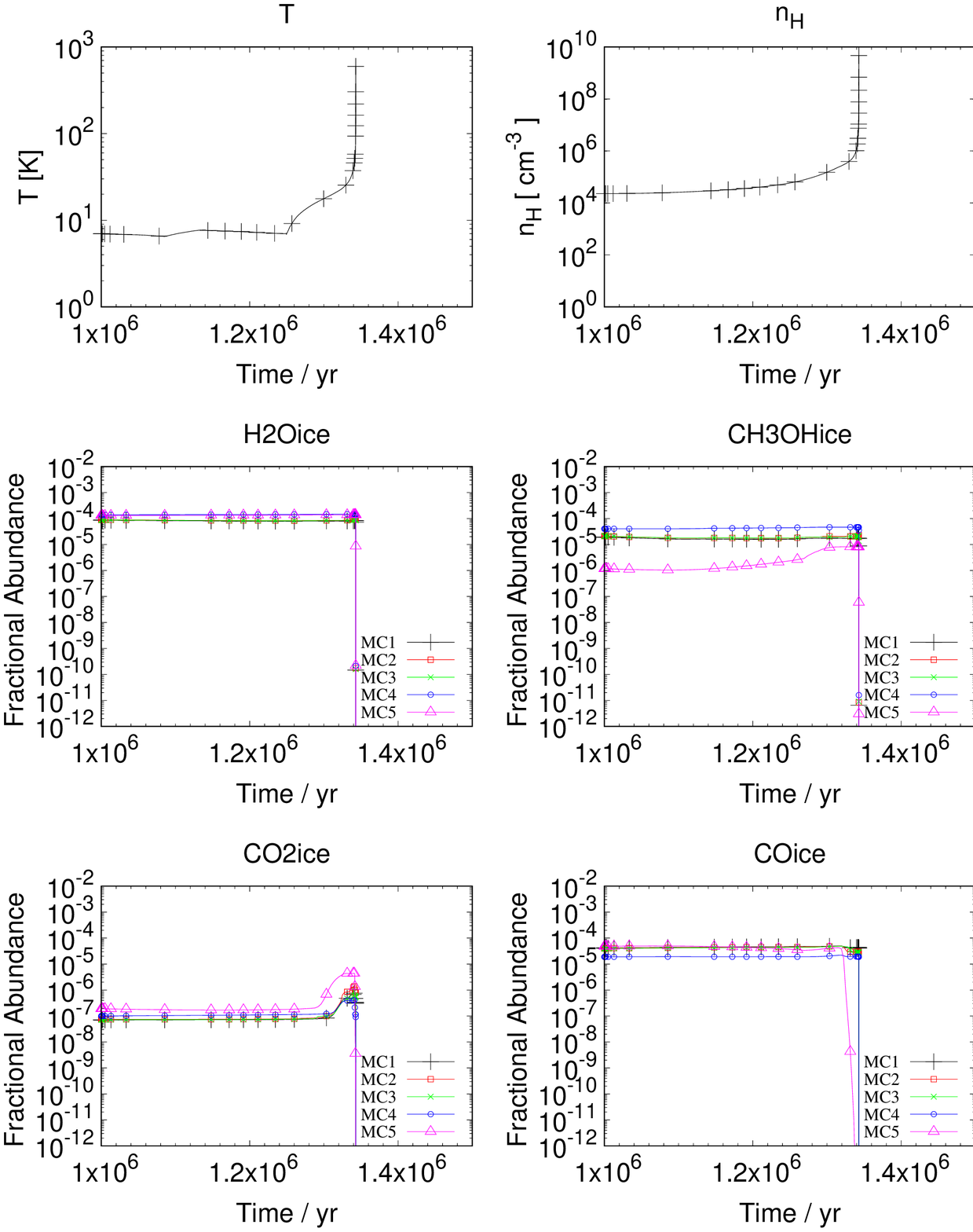}}
\caption{
The temporal variation of major granular species, JH$_2$O, JCO$_2$, JCO and JCH$_3$OH in the fluid parcel that reaches r=2.5 AU 
at $t_{\textrm{final}}$ 
after the molecular clouds start to contract in models MC1, MC2, MC3, MC4 and MC5.
The temporal evolution of the temperature and density for the fluid parcel is also plotted.
}
\label{Fig8}
\end{figure}

Fig.~\ref{Fig8} shows the temporal evolution of major granular species, temperature and density 
from the start of contraction to $t_{\textrm{final}}$ 
in the fluid particle that reaches r=2.5 AU in all models.
Significant amounts of products formed by photodissociation of water molecules cannot recombine to form water
again in models MC1, MC2 and MC3, which agrees with results calculated by more 
rigorous microscopic models~\citep{Chang2014} and molecular dynamic simulations~\citep{Andersson2008}. 
In the two-phase model MC5, products of the water photodissociation can easily recombine to form water again. 
In model MC4, because photons cannot penetrate into the ice mantle, water inside the ice mantle
cannot be photodissociated. Consequently, the abundances of JH$_2$O in models MC4 
and MC5 are higher than in models MC1, MC2 and MC3.  
Unlike water, the photodissociation products of granular methanol cannot easily recombine to 
form methanol again on dust grains in model MC5, so the methanol ice abundance in model MC5 is the lowest.
The abundance of granular methanol in model MC4 is the highest because methanol in the ice mantle cannot 
be photodissociated in this  model. In the new multiphase models, the exponential 
decay of photodissociation reactions in the ice mantle is well reproduced, and therefore granular methanol 
abundances in models MC1, MC2 and MC3 are higher than in model MC5. Granular carbon dioxide is mainly 
produced by surface reactions:
JO + JHCO $\rightarrow$ JCO$_2$ + JH and JCO + JOH $\rightarrow$ JCO$_2$ + JH. At around 10 K, because neither JCO nor JOH can
diffuse quickly on grain surfaces, JO + JHCO $\rightarrow$ JCO$_2$ + JH is the dominant reaction to produce JCO$_2$.
The photodissociation of JH$_2$CO or methanol contributes significantly to the production of JHCO. As with methanol, 
photodissociation of JH$_2$CO occurs more frequently in two-phase models than in the new and basic multiphase models. 
Therefore, the abundance of JCO$_2$ in model MC5 is higher than the other models.
Fig.~\ref{Fig8} also shows that the abundances of JCO initially decrease as temperatures of the
fluid parcel increase and then
remain almost constant until most water molecules sublime in models MC1, MC2 and MC3.
The abundance of JCO initially decreases because ICO diffuse into the active layers and then sublime.
After all ICO and gCO molecules sublime, significant amounts of KCO are still locked in the normal sites in 
models MC1, MC2 and MC3, thus the abundances of JCO remain almost constant until most water molecules sublime.
We calculate $p=N_{JCO}^{30K}/N_{JCO}^{20K}$ where $N_{JCO}^{30K}$ and $N_{JCO}^{20K}$ represent the population 
of JCO when the temperatures of the fluid parcel are 30 K and 20 K respectively.
We find $p=0.87,~ 0.60,~ 0.59$ and $0.90$ in models MC1, MC2, MC3 and MC4 respectively while almost
all JCO molecules sublime in model MC5 at 30 K.
Similarly,  ICO$_2$ and gCO$_2$ molecules sublime at lower temperatures and then KCO$_2$ sublime
when most water molecules sublime. However, the time interval between the sublimation of ICO$_2$ 
and KCO$_2$ is only about $10^3$ yr because the temperature of the fluid parcel increases quickly 
after ICO$_2$ and gCO$_2$ start to sublime. Thus, JCO$_2$ molecules  disappear almost at the same time in all models.

\subsection{Distribution of Molecules}

\begin{figure}
\centering
\resizebox{15cm}{15cm}{\includegraphics{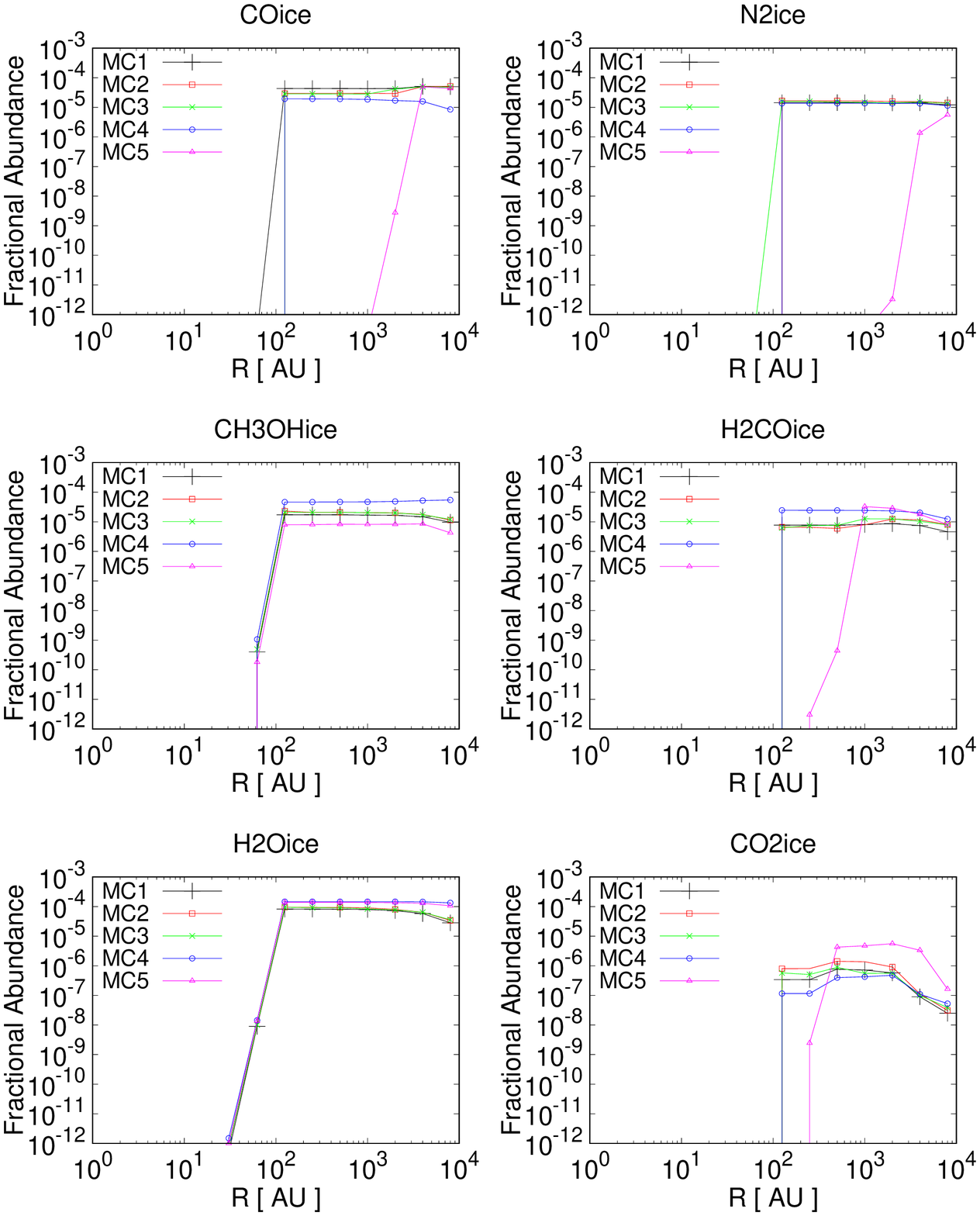}}
\caption{
Radial distributions of selected granular species in the protostellar core in different models at t$_{\textrm{final}}$.
}
\label{Fig9}
\end{figure}

\begin{figure}
\centering
\resizebox{15cm}{15cm}{\includegraphics{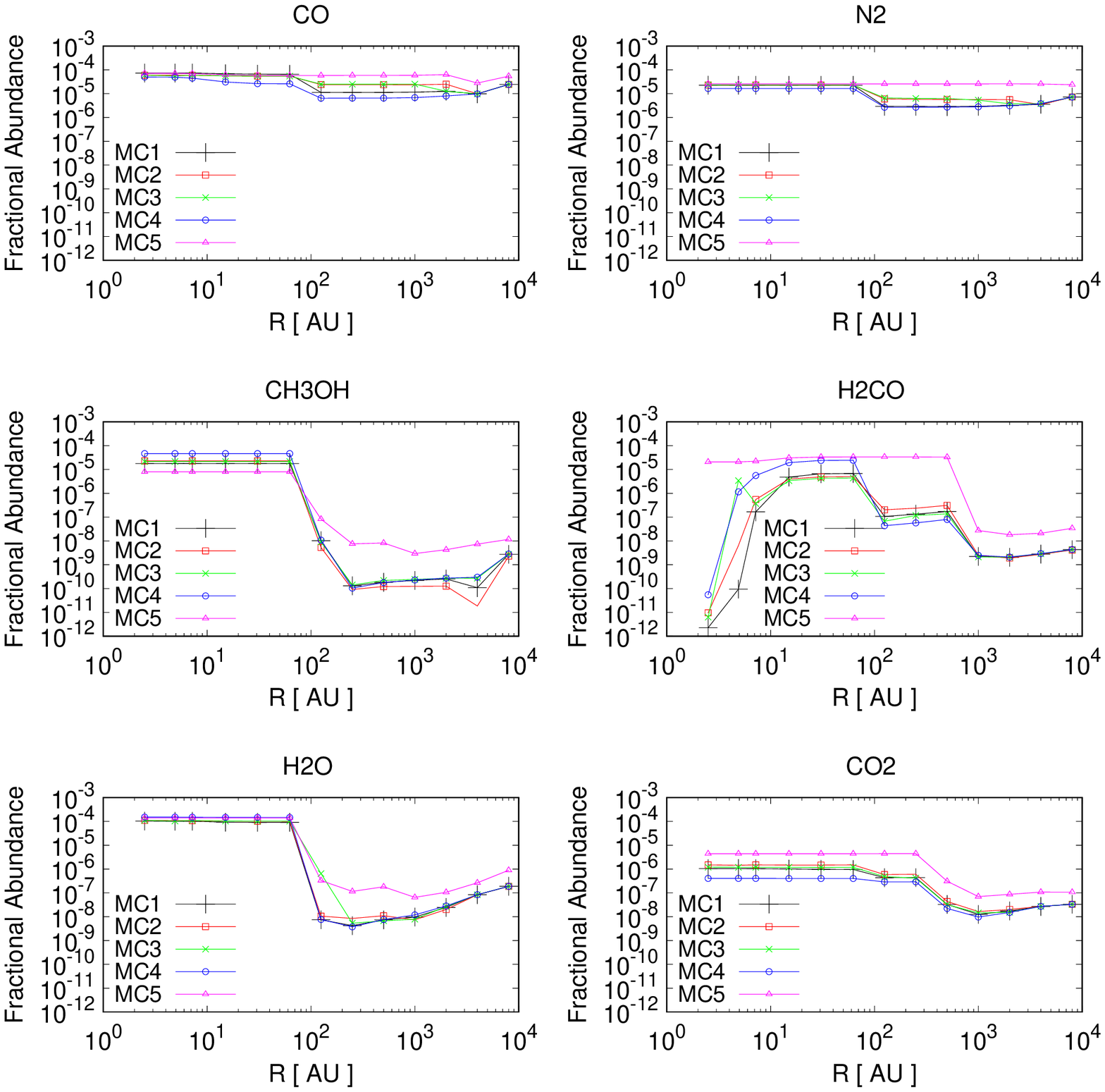}}
\caption{
Radial distributions of selected gas-phase species in the protostellar core in different models at t$_{\textrm{final}}$.
}
\label{Fig10}
\end{figure}

Fig. \ref{Fig9} and Fig. \ref{Fig10} show, respectively, the radial distributions of water, 
methanol, CO, H$_2$CO, CO$_2$ and N$_2$
on grain and in gas at $t_{\textrm{final}}$.  
The molecular evolution of 13 fluid parcels was calculated for each model in order to derive the radial distributions.
Fig. \ref{Fig9} shows that abundant water ice exists in the outer regions ($r>125$ AU) where the temperatures are below the sublimation 
temperature of water ice regardless of models.
The distribution of CO ice varies among model. In model MC5, most CO ice sublimes inside the radius of  
3000 AU. However, as discussed earlier, CO ice sublime via two processes in the basic and new multiphase models; 
significant amount of CO ice still exists between 3000 AU and 125 AU
because CO molecules locked in the normal sites cannot desorb until the water ice mantle sublimes. 
Interstitial CO molecules are able to desorb when the temperature is higher than $\sim~20$ K so that
JCO abundance drops slightly between 125 AU and 3000 AU in models MC2 and MC3.
The distribution of granular CO$_2$, H$_2$CO, and N$_2$ is 
similar to that of CO ice in each model. The desorption energy of methanol is close to that of water. 
Thus, methanol ice and water ice disappear in almost the same inner regions.

The abundances of gas phase water, methanol, CO, CO$_2$ and N$_2$ typically 
increase in all models in regions where these species sublime.
The radial distribution of gas phase H$_2$CO clearly shows that JH$_2$CO sublime via two processes.
The gas phase H$_2$CO abundance increases inward in two steps in models MC1, MC2, MC3 and MC4
because gH$_2$CO and IH$_2$CO molecules desorb first followed by the desorption of KH$_2$CO molecules.    
We can also see that the gas-phase H$_2$CO abundance typically drops in inner regions with $r<20$ AU 
because of the gas phase destruction of H$_2$CO. Moreover, gas-phase H$_2$CO abundance drops
inward more steeply in models MC1, MC2, MC3 and MC4 than in model MC5. Gas-phase H$_2$CO
can be destructed by the reaction with CH$_5$O$^{+}$, which 
can be produced by the reactions between gas-phase methanol and other ions.
The abundance of methanol ice in models MC1, MC2, MC3 and MC4 are much higher than in model MC5, so more H$_2$CO is destructed
in regions where the temperatures are high enough to desorb methanol ice.

\begin{figure}
\centering
\resizebox{10cm}{12cm}{\includegraphics{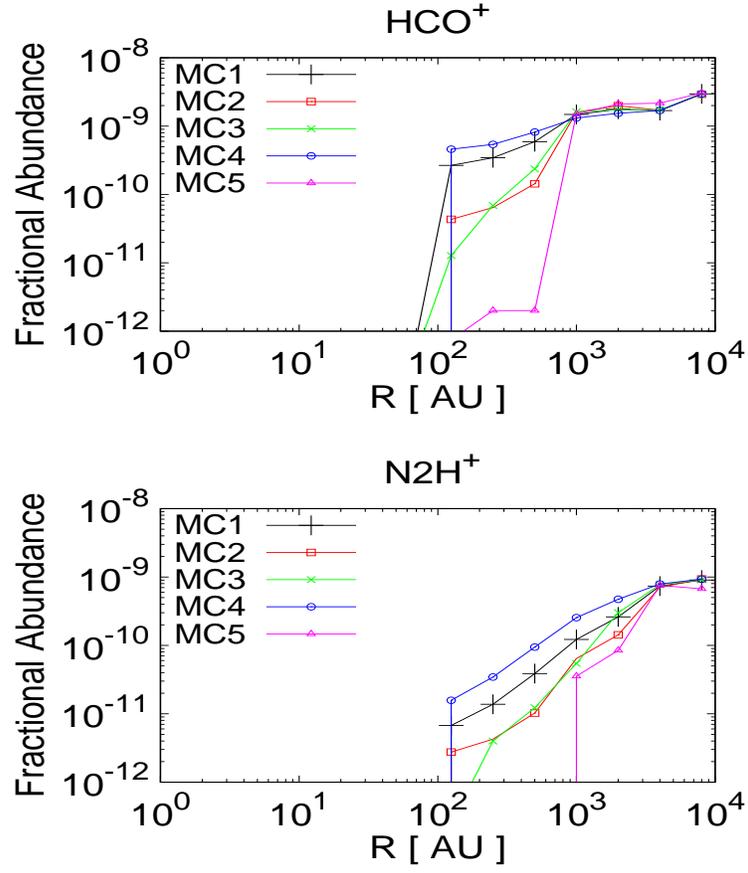}}
\caption{
Radial distributions of simple gas-species in the protostellar core of different models at t$_{final}$.
}
\label{Fig11}
\end{figure}

Fig.~\ref{Fig11} shows the radial distribution of gas phase N$_2$H$^{+}$ and HCO$^{+}$ abundances at $t_{\textrm{final}}$. 
The reaction H$_3^{+}$ + N$_2$ $\rightarrow$ N$_2$H$^{+}$ + H$_2$ is 
the major formation route of N$_2$H$^{+}$ while its destruction route  
is  N$_2$H$^{+}$ + CO $\rightarrow$ HCO$^{+}$ + N$_2$. So CO sublimation can decrease the abundance of N$_2$H$^{+}$
while N$_2$ sublimation increases its abundance.
We can see that the abundance of N$_2$H$^{+}$
drops inward in regions where both JN$_2$ ice and JCO ice sublime from grain surfaces in model MC5. 
Because a significant amount of KCO molecules are locked in the ice mantles in models MC1, MC2, MC3 and MC4,
N$_2$H$^{+}$ abundance drops inward much less steeply. 
The main destruction pathways for HCO$^{+}$ are its reactions  
with HCN or H$_2$CO after these two molecules sublime.
Since significant amounts of KH$_2$CO and KHCN are locked in the ice bulk in models MC1, MC2, MC3 and MC4,
HCO$^{+}$ abundance also drops much less steeply in these four models than in model MC5.

\begin{figure}
\centering
\resizebox{12cm}{12cm}{\includegraphics{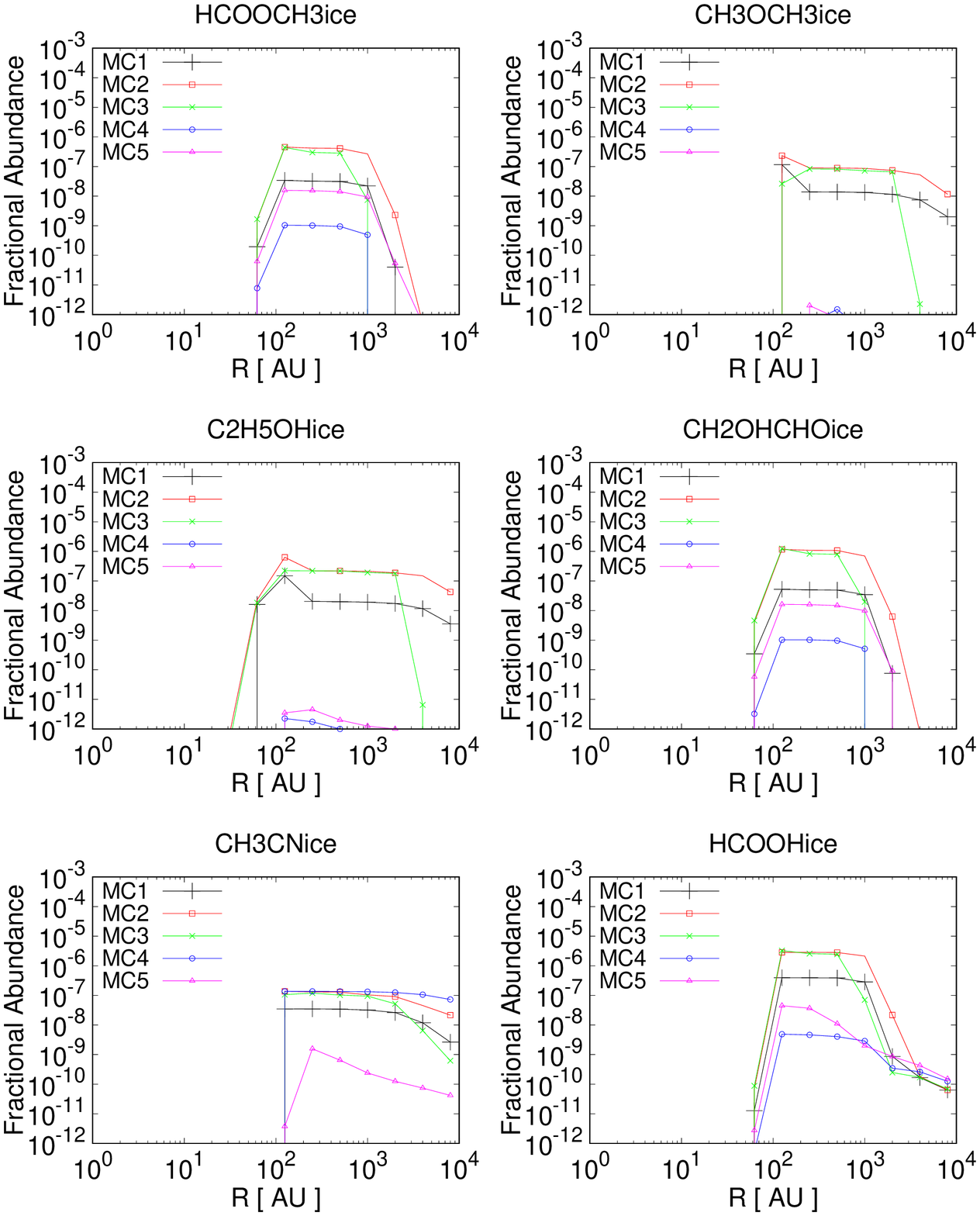}}
\caption{
Radial distributions of selected granular COM abundances at $t_{\textrm{final}}$.
}
\label{Fig12}
\end{figure}

Fig.~\ref{Fig12} shows the radial distributions of selected granular COM abundances at $t_{\textrm{final}}$. 
If the desorption energy of a COM is larger than that of water ice, the COM can sublime only 
in regions where the temperature is high enough to desorb the COM. However, if the 
desorption energy of the COM is smaller than that of water ice,
the COM sublime in two different regions as discussed earlier.
The COM abundances vary in outer regions in different models. The CH$_3$OCH$_3$ 
and C$_2$H$_5$OH ices extend to 8000 AU because JCH$_3$ can diffuse in ice mantles to form these species  
at around 8000 AU where the temperature is sightly above 16 K in models MC1 and MC2.  
Because the bulk diffusion barriers of interstitial radicals in model MC3 are larger than 
those in model MC1 and MC2, JCH$_3$OCH$_3$ and JC$_2$H$_5$OH are formed in warmer regions (~3000 AU) in model MC3.
Little JCH$_3$OCH$_3$ and JC$_2$H$_5$OH are produced 
in models MC4 and MC5, so JCH$_3$OCH$_3$ and JC$_2$H$_5$OH only exist in a narrow range.
Similarly, we can explain the distributions of HCOOCH$_3$ ice and CH$_2$OHCHO ice. 
Because CH$_3$CN ice is 
mainly produced by surface hydrogenation reactions at around 10 K in the basic and new multiphase models, 
CH$_3$CN ice extends to 8000 AU in models MC1, MC2, MC3 and MC4. However, in the two-phase model MC5, 
the CH$_3$CN ice abundance at r=8000 AU is much lower than that in the other models 
because of the absence of the exponential decay with depth of photodissociation for CH$_3$CN ice. 
The peak of the CH$_3$CN ice abundance occurs around 250 AU where Av is large enough to attenuate FUV radiation so that
only a small amount of CH$_3$CN ice can be photodissociated while CH$_3$CN ice can also be efficiently formed 
by the recombination of JCN and JCH$_3$~\citep{Garrod2008}.
A small amount of HCOOH ice also exists around 8000 AU because HCOOH can be produced in the gas phase 
and deposited on dust grains.

\begin{figure}
\centering
\resizebox{12cm}{12cm}{\includegraphics{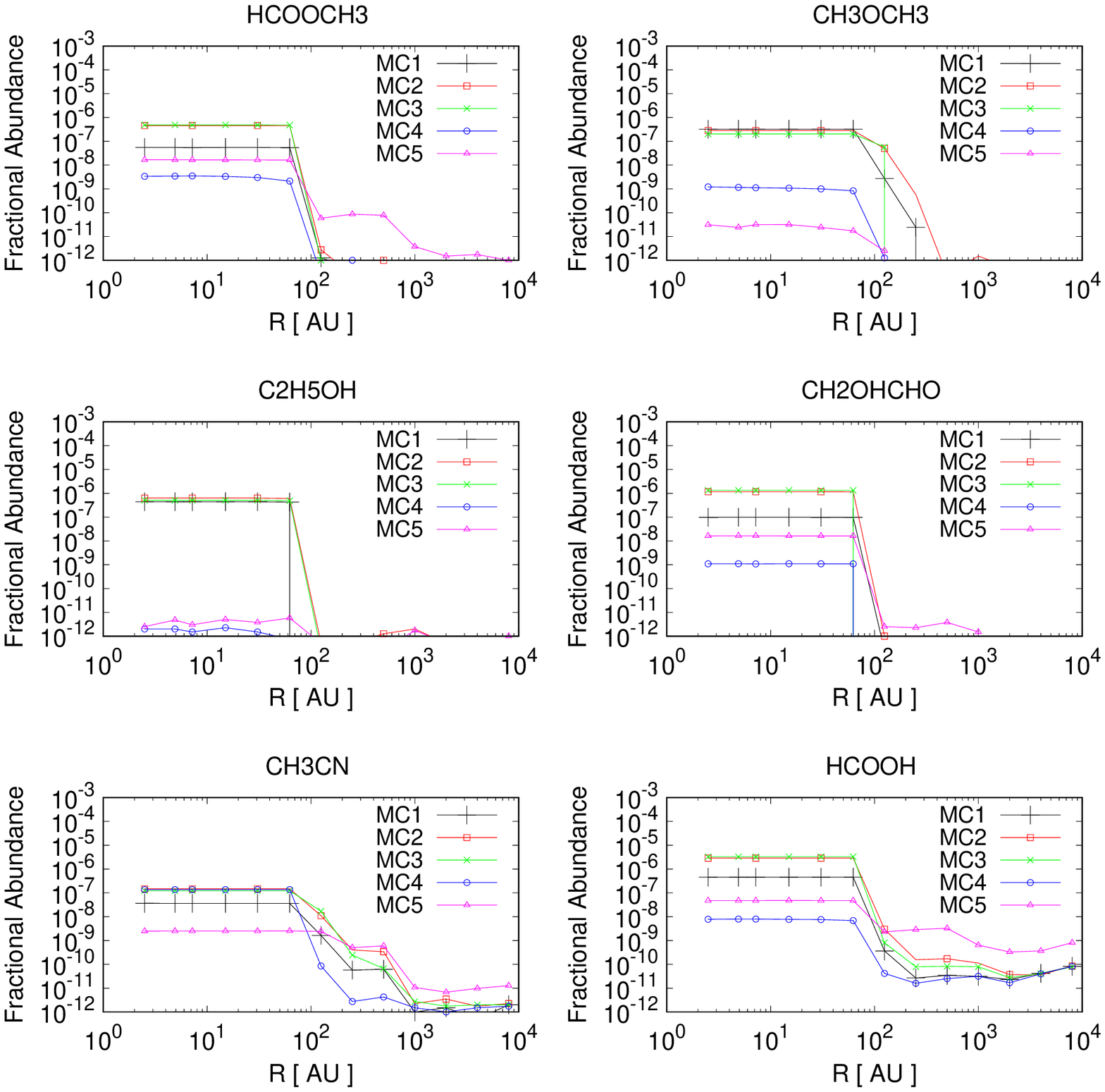}}
\caption{
Radial distributions of selected gas phase COM abundances at $t_{\textrm{final}}$.
}
\label{Fig13}
\end{figure}

The radial distributions of selected gas phase COMs abundances at $t_{\textrm{final}}$ are shown in Fig.~\ref{Fig13}.
The abundances of the COMs significantly increase in the inner regions where the ice mantle disappears 
for two reasons. First, the sublimation of COMs that are
mainly formed on dust grains can increase their abundances in gas phase. Second, ice mantle species other than COMs 
also produce COMs in the gas phase when they sublime.

\begin{figure}
\centering
\resizebox{12cm}{12cm}{\includegraphics{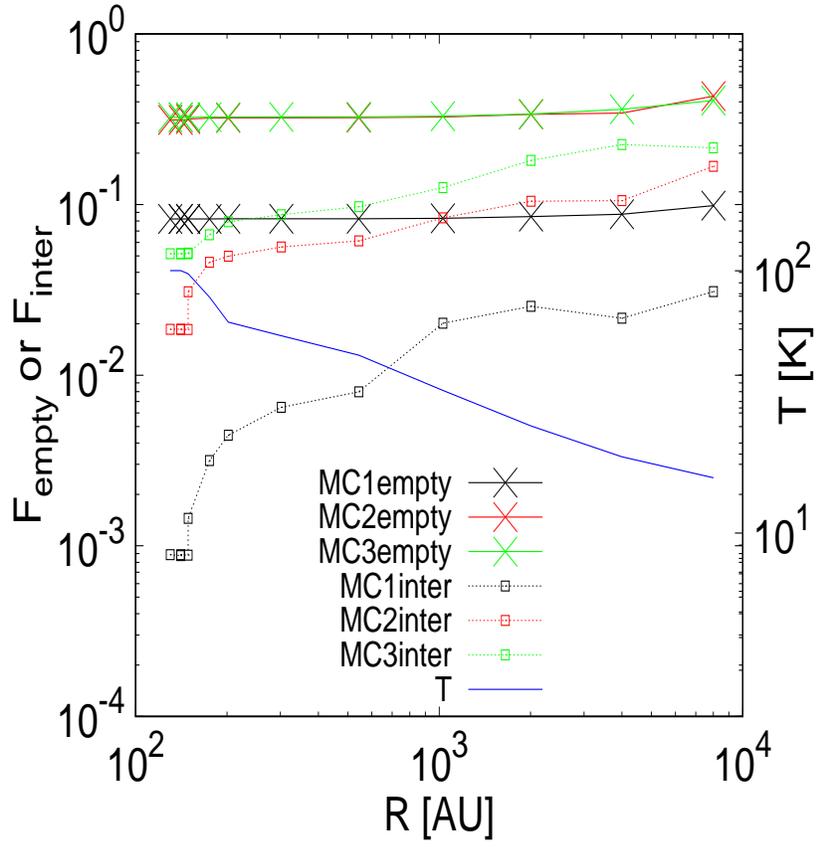}}
\caption{ 
Radial distributions of F$_{\textrm{empty}}$ and F$_{\textrm{inter}}$ predicted by the new multiphase models 
at t= t2 + $9.279 \times 10^{4}$ yr.
MC1empty, MC2empty and MC3empty represent F$_{\textrm{empty}}$ in models MC1, MC2 and MC3 respectively
while MC1inter, MC2inter and MC3inter represent F$_{\textrm{inter}}$ 
in models MC1, MC2 and MC3 respectively. 
}
\label{Fig14}
\end{figure}

Finally, Fig.~\ref{Fig14} shows the radial distributions of the fraction of empty normal sites 
(F$_{\textrm{empty}}$) and the fraction of interstitial species  
(F$_{\textrm{inter}}$) predicted by the new multiphase models at t = t2 + $9.279 \times 10^{4}$ yr
when the temperature of the inner most fluid parcel is 100 K. 
The fraction of empty normal sites is defined as, 
$F_{\textrm{empty}} = {\textrm{Total}}_{\textrm{empty}}/(N_L N_s) $
where ${\textrm{Total}}_{\textrm{empty}}$ is the population of empty normal sites in the bulk of ice.
Similarly, $F_{\textrm{inter}} = {\textrm{Total}}_{\textrm{inter}}/N_L N_s $, where
${\textrm{Total}}_{\textrm{inter}}$ is the population of all interstitial species in the bulk of ice.
Because more photodissociation fragments occupy interstitial binding sites in models with larger $\alpha$ value,
both F$_{\textrm{empty}}$ and F$_{\textrm{inter}}$ are larger in models MC2 and MC3 than 
in model MC1. On the other hand, because more interstitial species can diffuse into the active layers or participate reactions 
inside the bulk of ice as the temperature increases,
F$_{\textrm{inter}}$ drops quickly inward in all models.
On the other hand, F$_{\textrm{empty}}$ and F$_{\textrm{inter}}$ typically are
slightly larger in outer regions in all models because of stronger radiation in these regions.

\section{Comparison With Observations and Previous Models}\label{comp}

We compare our theoretical results with observations and 
the rate equation two-phase (RETP) model~\citep{Aikawa2008} results in this section. 
Most COMs were found within a few arcseconds from the
center of the core~\citep{Kuan2004}. On the other hand, most gaseous COM abundances 
do not vary much at $r\leq 100$ AU in our model results. Therefore, as in \citet{Aikawa2008},
the abundances of gaseous COMs in the fluid parcel that reaches 30.6 AU at $t_{\textrm{final}}$ 
are compared with the gaseous observations. The temperature and density of hydrogen nuclei of the fluid parcel are about
198 K  and 1.49$\times 10^8$ cm$^{-3}$, respectively at $t_{\textrm{final}}$. 
The granular species exist at outer regions where water ice is abundant ($r>100$ AU). 
Therefore, we calculate the average abundances of granular species relative to water ice in outer 7 fluid parcels 
in which water ice cannot sublime at $t_{\textrm{final}}$ to compare with observations of ices.
The average abundance of granular species Ji relative to water ice, $\bar{N}(Ji)$, is calculated as
$\bar{N}(Ji) = \frac{\sum_{m=1}^{m=7} N(Ji)_m}{\sum_{m=1}^{m=7} N(JH_2O)_m}$,
where $N(Ji)_m$ and $N(JH_2O)_m$ are the population of species Ji and JH$_2$O in m-th fluid parcel respectively.
The first fluid parcel is the outermost one and m increases as the distance from the fluid parcel to the
center of the protostar decreases.

\citet{Bergner2017} recently observed 16 low-mass protostars using the IRAM 30 m telescope, 
 and reported the median abundances of a few COMs.  \citet{Bergner2017} did not observe the abundance of glycolaldehyde, 
but its abundance toward low mass protostar IRAS 16293-2422 was recently reported by\citet{Jorgensen2016}.
Table~\ref{table33} summarizes
the median COM abundances with respect to methanol 
derived by the observations and our models.
Overall, the new multiphase models agree with observations  better than
the basic multiphase model and two phase models. 
Model MC1 is the only model that can reproduce the observed abundances of all COMs 
within one order of magnitude. 
All COM abundances other than that of glycolaldehyde are well predicted by models MC2 and MC3
while models MC4 severely underestimate the abundances of COMs other than CH$_3$CN.
Models MC5 and RETP are both simple two-phase models, but COM abundances predicted by model RETP can
be more than one order of magnitude higher than that by model MC5. 
One possible reason is the different reaction networks used in simulations.
The discrepancy might also be explained by the finite size effect~\citep{Vasyunin2009}.
Comparing with observations, both models severely underestimate the abundances of HCOOCH$_3$ and CH$_3$OCH$_3$.

\begin{table}
\caption{Comparison of Observational COM abundances with Theoretical Results}
\label{table33}
\begin{tabular}{cccccc}
  \hline \hline
   Model          & $^b$HCOOCH$_3$    & $^b$CH$_3$OCH$_3$     & $^b$CH$_3$CN & $^b$CH$_3$CHO & $^c$CH$_2$OHCHO\\
  \hline
   Observations   &  3.1(-2)        & 5.3(-2)          &  4.8(-3)           & 4.4(-2)  & 3.4(-3) \\ 
  MC1            &  3.1(-3)         &  1.8(-2)          &  2.0(-3)          & 1.3(-2)  & 5.4(-3) \\    

  MC2            &  2.0(-2)        &  1.2(-2)         &  6.5(-3)            &  1.7(-2)  & {\bf 5.0(-2)} \\

  MC3            &  2.2(-2)        &  9.1(-3)         &  5.5(-3)            & 1.8(-2)   & {\bf 6.1(-2)}  \\

  MC4            &  {\bf 6.5(-5)}  &  {\bf 2.2(-5)}    &  3.0(-3)           & {\bf 6.7(-7)}   & {\bf 2.4(-5)}  \\

  MC5            &  {\bf 2.0(-3)}  &  {\bf 3.0(-6)}    &  {\bf 3.1(-4)}     & {\bf 1.5(-4)}  & 2.0(-3)\\

  RETP$^{a}$       &  {\bf 6.0(-4)}  &  {\bf 1.2(-4)}  &  1.0(-2) &        & \\
  \hline
\end{tabular}
\medskip{\protect\\
Notes.\protect\\
${a}$ \citet{Aikawa2008}.\\
${b}$ Observational results are median values toward 16 low-mass protostars~\citep{Bergner2017}.\\ 
${c}$ The observed abundance of CH$_2$OHCHO is toward IRAS 16293-2422~\citep{Jorgensen2016}.\\
All abundances are with respect to methanol.
a(-b) means a$\times 10^{-b}$. Boldface indicates more than one order of magnitude disagreement
between model and observations. 
}
\end{table}

Table~\ref{table4} shows major granular species abundances predicted by different models and the 
observed ice abundances toward low-mass protostars by~\citet{Oberg2011}. All the abundances are expressed in percentages 
with respect to the water ice. We can see that the abundances of CO$_2$ are much lower than 
the observed value. Moreover, the CO$_2$ abundance in the model MC5 is the highest. 
The reaction CO + OH $\rightarrow$ CO$_2$ + H is the most efficient reaction to convert CO to CO$_2$ on grain surfaces
when the temperature is higher than 12 K~\citep{Garrod2013}.
However, most granular species are formed when the temperature of the core is around 10 K, 
so the conversion from CO to CO$_2$ on grain surfaces is not efficient in our models.
The chain reaction mechanism ~\citep{Chang2007, Chang2012} allows surface OH to react with 
surface CO immediately after surface OH is 
produced in ice. So, if it is introduced in surface and bulk models, 
the reaction CO + OH $\rightarrow$ CO$_2$ + H can proceed efficiently at around 10 K. However, rigorous
treatment of the chain reaction mechanism requires the microscopic Monte Carlo method, which is much more 
computationally expensive.
On the other hand, the chain-reaction mechanism can be approximately 
implemented by the rate equation approach~\citep{Garrod2011}.
Following \citet{Garrod2011}, we implement the chain-reaction mechanism in models MC1, MC2 and MC3. 
The abundances of CO$_2$ ice increase to more than 10\% relative to water ice with 
the chain-reaction mechanism.

The abundance of JCO predicted by model MC5 is the lowest among all models because JCO cannot be trapped in ice mantle
in the two-phase model.
We can also see that granular methanol abundances are 
overestimated in the basic and new multiphase models. Because of the exponential decay of the rate 
of photodissociation reactions with depth in the new multiphase models, methanol molecules 
are less likely to be photodissociated in models MC1, MC2 and MC3 
than in model MC5. Therefore, granular methanol abundances in models MC1, MC2 and MC3 
are higher than that in model MC5.
Moreover, the ice mantle is completely inert in model MC4, thus the 
methanol ice abundance in model MC4 is even higher than that in 
the new multiphase models.  \citet{Pontoppidan2003} found that the methanol ice abundances are
more than 15\% relative to water ice toward 3 low-mass protostars, which agrees with our new multiphase model 
results. The different formation history of protostars, for instance the time scale before collapse, 
may be able to explain the discrepancy.   
Methane and NH$_3$ abundances predicted by theoretical models agree reasonably well with observations. 

 \begin{table}
 \caption{Comparison of Model and Observational Granular Species Abundances}
\label{table4}
 \begin{tabular}{c c c c c c c}
    \hline
    Model    &JCO    &JCO$_2$ & JH$_2$CO   &JCH$_3$OH &JCH$_4$ & JNH$_3$\\
   \hline
    \multirow{1}{*}{MC1}   &65.5   &0.59   &10.8  &22.6 &7.2  & 11.4 \\
          \hline              
    \multirow{1}{*}{MC2}   &45.3  &0.99   &10.7  &24.4 &6.8 & 11.7\\
          \hline              
                        
    \multirow{1}{*}{MC3}   &46.9   &0.60   &12.1  &24.6 &6.9 & 11.6\\
                        
           \hline             
    \multirow{1}{*}{MC4}   &11.6   &0.17   &15.2  &33.8 &7.6 & 25.0 \\  
                        
             \hline    

   \multirow{1}{*}{MC5}    &10.0  &2.0   &9.4   &5.8 &1.1 & 13.4 \\
  
   \hline
    \multirow{1}{*}{Observation }  &29  &29  & --   & 3 & 5 & 5 \\  
                        
                \hline                                       
    \end{tabular}
\tablecomments{The abundances of major ice mantle components are percentages with respect to the water ice.
 Observational results are from ~\citet{Oberg2011}. }
    \end{table}

\begin{figure}
\centering
\resizebox{12cm}{12cm}{\includegraphics{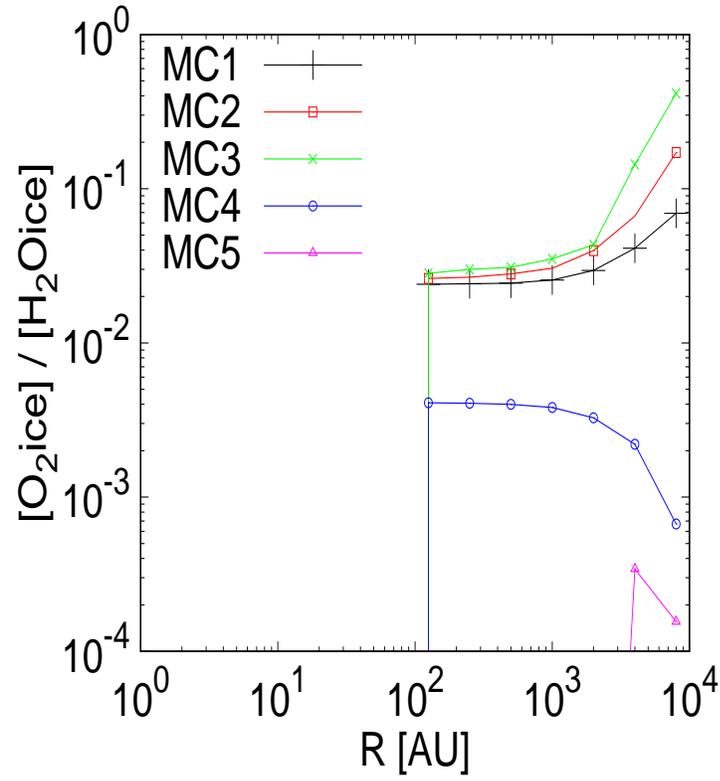}}
\caption{
Radial distributions of O$_2$ ice abundances at $t_{\textrm{final}}$.
The abundances are relative to water ice.
}
\label{Fig15}
\end{figure}

It was recently found that there are abundant O$_2$ molecules (O$_2$/H$_2$O in the range of 1\%-10\%,  mean value $\sim$ 4\%) 
in the coma of comet 67P/Churyumov-Gerasimenko(67/C-G)~\citep{Bieler2015}.
\citet{Mousis2016} argued that O$_2$ can be formed radiolysis of water ice in parental molecular clouds 
and then trapped in clathrates in the solar nebula stage. \citet{Taquet2016}, on the other hand, solved
detailed gas-grain reaction networks and showed that primordial (i.e. interstellar) O$_2$ ice in dark molecular clouds can 
be as abundant as observed in 67/C-G, if the dark clouds are rather dense and warm. However,
it was found that the JO$_2$/JH$_2$O ratio is well below 0.01 in most fluid parcels during 
the formation of protostellar disks and the high JO$_2$/JH$_2$O ratio only exists 
in the upper layers of disks\citep{Taquet2016}.
Since the bulk ice mantle is partially active in our models while it is completely inert
in \citet{Taquet2016}, it is worth presenting how O$_2$ ice evolve in our new multiphase models.
Figure~\ref{Fig15} shows the
radial distribution of O$_2$ ice relative to water ice at $t_{\textrm{final}}$. We can see that our new multiphase
models predict that the abundance of O$_2$ ice is around 10\% of water ice in the outermost fluid parcel,
whose temperature is around 16 K and is about 8000 AU from the center of the protostar.
The abundance of O$_2$ ice drops inward to around 3\% of water ice abundance at r=125 AU. 
Although our results cannot be directly compared with that of~\citet{Taquet2016}, because our core model is spherical,
relatively abundant O$_2$ ice survives down to the water sublimation radius in our new multiphase models.
On the other hand, the abundances of O$_2$ ice in models MC4 and MC5 are much lower than 4\%. 

In recent years, COMs are detected also in prestellar cores~\citep{Bacmann2012}. 
It motivated intense studies on grain-surface reactions, non-thermal desorption, 
and gas-phase formation of COMs at cold temperatures ($\sim$ 10 K)~\citep{Vasyuninherbst2013,Chang2016,Vasyunin2017,Chen2018}.
It would thus be useful to compare our model results before the gravitational contraction with the observations of prestellar cores. 
The fluid parcels in our model are, however, located at the radius of $\sim 10^4$ AU before the gravitational contraction, 
so that they fall to the central region of the protostellar core. We thus compare our gas-phase abundances before the contraction 
with the observations of dense cloud TMC-1, rather than prestellar cores.
The observed fractional abundances of gas-phase H$_2$CO and CH$_3$OH toward the dense
cloud TMC-1 are $5\times 10^{-8}$ and $3\times 10^{-9}$ respectively~\citep{Smith2004}.
On the other hand, \citet{Soma2018} recently found that H$_2$CO and CH$_3$OH abundances could be similar.
Fig. \ref{Fig5} shows that models MC1-4 either underestimate H$_2$CO or CH$_3$OH abundances while the 
two-phase model can reproduce the observed abundances of H$_2$CO and CH$_3$OH well at the time later than $1\times 10^5$ yr.
So, more efficient desorption mechanism such as reactive desorption should be introduced into models MC1-4 in order to reproduce
the observed abundances of CH$_3$OH toward TMC-1~\citep{Garrod2007}.
On the other hand, models MC1-4 can predict that abundances of gas-phase H$_2$CO and CH$_3$OH
are similar at the time $1\times 10^6$ yr 
while difference of H$_2$CO and CH$_3$OH abundances in the two-phase model is usually around one order of magnitude.

The visual extinction of fluid parcels in our model are around 4 before contraction, which is much lower
than that under standard physical conditions (Av=10).
FUV photons can not only desorb molecules out of grains, 
but also destruct these molecules in gas phase and ice mantle.  
We perform test simulations of all chemical models using the standard physical conditions in
order to see if the lower visual extinction makes any difference to the gas-phase H$_2$CO and CH$_3$OH abundances.
In the test simulations, the visual extinction is set to be 10, the density of H nuclei is $2\times 10^4$ cm$^{-3}$ 
while the temperature is 10 K.
We found that the abundances of H$_2$CO and CH$_3$OH at the time $2\times 10^5$ yr using the standard physical conditions 
and in the fluid parcel differ by less than a factor of two in all models.
At the time $1\times 10^6$ yr, a factor of two or less difference is introduced into 
the abundances of these two species in models MC1-4 and H$_2$CO abundance in the two-phase model
by the lower visual extinction.
However, at the later time, CH$_3$OH and H$_2$CO abundances in the two-phase model using the standard physical conditions are similar,
which is more than one order of magnitude larger than CH$_3$OH abundance in the fluid parcel.

\section{Conclusions and Discussions}\label{diss}

In this work, we study gas-grain chemistry in a collapsing core during star formation 
using a new multiphase astrochemical approach. 
Our emphasis is on the formation of complex organic molecules. We also pay attention to the gas-ice balance of CO. 
The physical structure of the collapsing core is based on a 1-D RHD model. 
Our new multiphase model results are compared with older two-phase and basic multiphase results. The abundances of COMs 
in our new multiphase models are found to be more than one order of magnitude higher than in two-phase models.
This  difference can be explained by two reasons.
First, because of the exponential decay with depth of photodissociation reaction rates, 
COMs buried in the ice mantle in the new multiphase 
models are less likely to be photodissociated than in the two-phase model. 
Second, abundant radicals accumulate in the ice mantle
when the temperature is around 10 K in the new multiphase models. 
The accumulated radicals diffuse inside ice mantle and react
with each other to form COMs when the temperature increases. 
In the basic multiphase model, the abundances of COMs such as methanol, which are mainly formed by hydrogenation 
reactions at 10 K, are higher than in the new multiphase 
models. However, 
the abundances of COMs, that are mainly formed by radical 
recombination at higher temperatures, are much lower in the basic multiphase model 
than in the new multiphase models because radicals cannot accumulate in the bulk of 
ice mantle in the basic multiphase model. 
The desorption of CO in the new multiphase models and in the two-phase model is different. 
Because significant amounts of CO are locked in the ice bulk mantle, solid CO molecules sublime via two processes
in the new and basic multiphase model. First, solid CO occupying interstitial binding sites or in the active 
layers desorbs when the temperature reaches the sublimation points. Second, solid CO sublimes together 
with water ice when the temperature is high enough to desorb water ice. 
The amounts of CO locked in the ice mantles depend on the 
$\alpha$ value, which is a parameter that determines the probability 
photofragments enter normal sites to be locked in the ice mantle. 
The larger the $\alpha$ value is, the fewer CO molecules are locked in the 
ice mantle. Because dust grains are covered with ice mantles at $T<100$ K, 
significant amounts of solid CO still exist at $T<100$ K.    
 
We introduce a new COM formation scenario applicable during star formation. 
COMs are believed to be formed when the temperature 
of a collapsing core increases and radicals, which are generated by 
photodissociation reactions, recombine with 
each other~\citep{Garrod2006,Aikawa2008}. 
Almost all radicals that form COMs in the two-phase model were generated
when the temperature of the core increases~\citep{Garrod2006}.
In our multiphase model, most radicals that recombine to form COMs have 
a primordial origin when the temperature is around 10 K before the core starts to collapse.
The abundances of COMs such as CH$_3$OCH$_3$ in the two-phase model are much lower than observations, 
which suggests that more radicals need to be generated in ice so that these radicals can recombine to form more COMs.
In our new multiphase model, radicals formed in the prestellar phase can be stored, thus more COMs can be
formed upon warm-up.
Our COM formation scenario is also consistent with laboratory experiments, at least qualitatively. 
For example, \citet{Butscher2016} showed that abundant free radicals can be stored in ice 
mantles by photolysis of water rich ice mixture at around 13 K while COMs can be 
formed by radical recombination when the temperature increases.   
Moreover, it was found that abundant free radicals can be frozen in the ice mantle 
which is photolysed by UV radiation at around 10 K~\citep{Schutte1991}. 
The recombination of free radicals can even lead to chemical explosion   
if there are enough free radicals in the ice mantle~\citep{Schutte1991}.

Our new multiphase models are also consistent with laboratory experiments of 
CO desorption~\citep{Collings2003, Sandford1988}. A fraction of CO is desorbed around 20 K and the rest 
is locked in water ice and thus desorbed together with water ice at $\gtrsim$ 100 K. 
The abundant O$_2$ found in the coma of comets 
must also be locked in ice if these O$_2$ molecules have a primordial origin,
because unlocked O$_2$ sublime at a much lower temperature.  
Our new multiphase models keep the O$_2$/water ratio as high as $\gtrsim$ 3\%, 
around the water ice sublimation region.   

It is interesting that COM formation and CO desorption, which are two independent questions, are linked together in our 
new multiphase models. The population of species unlocked in ice mantles, which can diffuse inside ice mantles before water ice 
sublime, play a central role in answering these two questions. 

Finally, we discuss the limitations of our modeling. First, the gas and dust temperatures are assumed equal. 
However, the gas densities of fluid parcel beyond 3000 AU are low enough for the gas 
and dust temperatures to be decoupled. 
In order to find out the effect of temperature decoupling on the chemical evolution of fluid parcels, 
we simulate a test model in which the dust temperatures are 2 K lower than the gas temperatures
when the gas temperatures are higher than 12 K.
We still assume that the gas temperatures and dust temperatures are equal when the gas temperatures are lower than 12 K
because surface chemistry on dust grains is not dependent on temperatures 
when the dust temperatures are between 8 K and 12 K~\citep{Garrod2011}.
The chemical model used in the test model is the same as that in model MC2. We only simulate the 
molecular evolution of the fluid parcel 
that reaches 4000 AU and 8000 AU (the outermost fluid parcel) at t$_{\textrm{final}}$. 
We found that the test model and model MC2 predict similar
abundances of all species in the fluid parcel that reaches 4000 AU at t$_{\textrm{final}}$. 
Moreover, other than granular COMs that are formed by the recombination of two radicals,
species abundances in the outermost fluid parcel predicted by the test model are also similar to these by model MC2.  
However, the abundances of granular COMs which are formed by the recombination of two radicals
in the test model may be much lower than in model MC2 because the dust temperatures in the test model are not high
enough so that radicals cannot diffuse and recombine to form COMs in ice.
For instance, model MC2 predicts that 
the fractional abundance of JCH$_3$OCH$_3$ in the outermost fluid at t$_{\textrm{final}}$ is around $10^{-8}$ while
JCH$_3$OCH$_3$ molecules can hardly be formed in the same fluid parcel in the test model because of the lower dust temperatures. 
Moreover, because the temperatures of inner fluid parcels  (r $<$ 4000 AU) are not lower than the fluid parcel that 
reaches 4000 AU at t$_{\textrm{final}}$, the test model and model MC2 should predict similar abundances of all species
in these inner fluid parcels.      

Another issue is the binding energy of atomic oxygen. In this work, the binding energy of atomic oxygen is set to be 800 K,
which is commonly used in astrochemical modeling. However, \citet{He2015} recently suggested a much higher value, 1660 K. In order to
investigate how the higher binding energy of atomic oxygen affect our simulation results, we run a test model in which 
the binding energy of atomic oxygen is set to be 1660 K. The chemical model used is model MC2. Hereafter, we call 
this test model TMC2. We simulate model TMC2 for all fluid parcels. We found that in the pre-collapse phase, 
the majority of species are not much affected by the binding energy of atomic oxygen, except for
the abundances of a few species whose production depends on the diffusion of O atoms.
At the time 10$^6$ yr, the abundance of ICO$_2$ in model TMC2 is more than one order 
of magnitude lower than model MC2 in all fluid parcels,
because the recombination of IO and IHCO is one of the major reaction to form ICO$_2$ at 10 K.
The abundance of interstitial O$_2$ is also reduced in model TMC2
because most IO$_2$ is formed by the recombination of IO and IO. 
At 10$^6$ yr, the abundances of IO$_2$ are around $2\times 10^{-6}$ in all fluid parcels in model MC2 
while few IO$_2$ molecules are produced in model TMC2. 
The abundances of JCO$_2$ and JO$_2$ also drop by about a factor of two and 
one order of magnitude respectively in model TMC2 in all fluid parcels.
At t$_{\textrm{final}}$, the abundances of all species predicted in model TMC2 
do not differ much from these in model MC2 in the inner fluid parcels (r $\leq$ 4000 AU).
However, the abundance of JO$_2$ in model TMC2 is
about two orders of magnitude lower than in MC2 in the outermost fluid parcels (T $\sim$ 16 K) at t$_{\textrm{final}}$.
Because the binding energy of atomic oxygen can strongly affect the formation of 
a few granular species such as JO$_2$ when the grain temperature is low (T $\leq$ 16 K), more work is needed to measure 
this binding energy more accurately.

Our new multiphase models have two strengths. First, the exponential decay of photodissociation 
reactions with depth into ice mantles is well mimicked. Second, species that are locked in the ice mantle and these
species that can diffuse in the ice mantle are well distinguished. 
To the best of our knowledge, 
few current astrochemical models have the first strength. 
On the other hand, the computational cost to simulate the new multiphase models is much lower than that required by 
the model introduced by \citet{Chang2014} because we use a macroscopic approach. 
We focus on the formation of COMs in this paper, but
our new multiphase models warrant further investigation. For example, the warm carbon-chain chemistry (WCCC)
is triggered by the sublimation of CH$_4$ from ice mantle~\citep{Sakai2008}, thus, the gas-ice balance of CH$_4$
is crucial for the WCCC. The gas-ice balance of volatile species in the new multiphase models
will be discussed in detail in a subsequent paper~\citep{Wang2018}.

Q. Chang is a research fellow of the One-Hundred-Talent project of the Chinese Academy of Sciences.
This work was funded by The National Natural Science foundation of China under grant 11673054.
YA wishes to acknowledge the support of the JSPS through KAKENHI Grant 16K13782, 16H00931 and 18H03718.
We thank Eric Herbst for a critical reading of the manuscript and helpful comments. 
We thank Ling Liao and Yao Wang for helpful discussions. 
We thank Fangfang Li and Long-Fei Chen for proofreading of the manuscript.
We thank our referee's constructive comments to improve the quality of the manuscript.
The Taurus High Performance Computing system 
of Xinjiang Astronomical Observatory was used for the simulations.

\end{document}